\documentclass[reprint,superscriptaddress,amsmath,amssymb,aps,prl]{revtex4-2}
\usepackage{graphicx} 
\usepackage{dcolumn}
\usepackage{bm}
\usepackage{natbib}
\usepackage[hidelinks]{hyperref}
\usepackage{xcolor}

\begin{document}

\title{On the relationship between equilibria and dynamics in large, random neuronal networks}

\author{Xiaoyu Yang}
\affiliation{Graduate Program in Physics and Astronomy, Stony Brook University, Stony Brook, NY, USA}
\affiliation{Department of Neurobiology and Behavior, Stony Brook University, Stony Brook, NY, USA}
\affiliation{Center for Neural Circuit Dynamics, Stony Brook University, Stony Brook, NY, USA}

\author{Giancarlo La Camera}
\email{giancarlo.lacamera@stonybrook.edu}
\affiliation{Department of Neurobiology and Behavior, Stony Brook University, Stony Brook, NY, USA}
\affiliation{Center for Neural Circuit Dynamics, Stony Brook University, Stony Brook, NY, USA}
\affiliation{AI Innovation Institute, Stony Brook University, Stony Brook, NY, USA}
\affiliation{Center for Advanced Computational Science, Stony Brook University, Stony Brook, NY, USA}
\affiliation{Graduate Programs in Neuroscience, Stony Brook University, Stony Brook, NY, USA}

\author{Gianluigi Mongillo}
\email{gianluigi.mongillo@gmail.com}
\affiliation{Sorbonne Universit\'e, INSERM, CNRS, Institut de la Vision, F-75012 Paris, France}
\affiliation{Centre National de la Recherche Scientifique, Paris, France}
\affiliation{School of Natural Sciences, Institute for Advanced Study, Princeton, NJ, USA.}

\begin{abstract}
    
We investigate the equilibria of a random model network exhibiting extensive chaos. In this regime, a large number of equilibria is present. They are all saddles with an extensive, but fractionally-small, number of unstable directions. Despite network's connectivity being completely random, the equilibria are strongly correlated and, as a result, they occupy a small region in the phase space. The attractor is inside this region. This geometry explains why the chaotic dynamics in these models can be described by a fractionally-small number of collective modes. 
\end{abstract}

\maketitle

Large systems of non-linear differential equations are a standard modeling framework for many natural and technological systems. Qualitative analysis, however, is of limited help in the study of their dynamics, because of the large number of degrees of freedom. One general approach, for suitably idealized systems, is dynamical mean-field theory (DMFT) \cite{crisanti2018path,clark2023dimension}. DMFT allows one to obtain a drastic, and exact, dimensionality reduction; one only needs to consider a single degree of freedom interacting with an effective `field', to be determined self-consistently. Unfortunately, though exact, the DMFT equations are usually still analytically intractable.

These difficulties have motivated the search for approaches complementary to DMFT. In particular, a promising approach focuses on determining the number of equilibria and their stability \cite{wainrib2013topological,ben2021counting,ros2023generalized,stubenrauch2025fixed,fournier2025non}. The hope is to use this information to infer qualitative/quantitative aspects of the dynamics, without solving the DMFT \cite{wainrib2013topological,stubenrauch2025fixed,fournier2025non}. Alternatively, this information could conceivably help in solving the DMFT numerically. 

In this letter, we focus on large, random neuronal networks as a model dynamical system. These networks exhibit a paradigmatic transition between a unique, stable equilibrium and a chaotic attractor \cite{sompolinsky1988chaos}. Their dynamics has been extensively studied and, hence, is fairly well understood \cite{sompolinsky1988chaos,crisanti2018path,clark2023dimension,engelken2023lyapunov}; in particular, the DMFT equations {\em can} be analytically solved in this case. Previous studies, however, did not investigate the spatial organization of the equilibria, and only one \cite{stubenrauch2025fixed} attempted to {\em locate} the dynamics with respect to the equilibria.

Our model provides an idealized description of the neural activity within a local cortical network (see \cite{si_link} for more details). It is as follows. The state of neuron $i$, $x_i$ $\left(i=1,\ldots,N\right)$, evolves according to
\begin{equation}\label{eq:dyn_x}
    \dot{x}_i=-x_i+\sqrt{N}+\frac{1}{\sqrt{N}}\sum_{j=1}^Nw_{ij}\phi_j\equiv v_i\left(\mathbf{x}\right),
\end{equation}

\noindent with $\phi_j\equiv\phi\left(x_j\right)$. The activation function, $\phi\left(x\right)$, determines neuron's firing rate as a function of its state. We take $\phi\left(x\right)\equiv x\Theta\left(x\right)$, where $\Theta\left(x\right)$ is the Heaviside function, such that $\phi_j$ is non-negative and does not saturate for large $x_j$ \cite{treves1990graded}. 

The weights, $w_{ij}$, are independent Gaussian deviates with $\mathbb{E}\left[w_{ij}\right]=-1$ and $\mathrm{Var}\left[w_{ij}\right]=\sigma^2_w$. Therefore, as in the cortex, recurrent connectivity is inhibition-dominated, because $\mathbb{E}\left[w_{ij}\right]<0$, and disordered, because $\mathrm{Var}\left[w_{ij}\right]>0$. The $1/\sqrt{N}$ scaling of the recurrent input (the sum over $j$) makes it $\mathcal{O}\left(\sqrt{N}\right)$ as the external, excitatory input ($+\sqrt{N}$). However, the network dynamics leads to steady states in which they cancel each other to leading order; therefore, our model network operates in the so-called balanced excitation-inhibition regime \cite{van1996chaos,renart2010asynchronous}, mimicking core features of cortical activity \cite{si_link}.

We investigate the equilibria of the network dynamics. The number of hyperbolic equilibria for a given sample $\mathbf{w}$, $P_{\mathbf{w}}$, is given by the so-called Kac-Rice formula \footnote{As an historical aside, we note that neither Kac nor Rice ever wrote such a formula, let alone with the purpose of counting the solutions of a large system of non-linear equations. Instead, for that precise purpose, Eq. \eqref{eq:kac_rice} and an accompanying theory first appear in \cite{bray1980metastable} to the best of our knowledge.}, i.e.,
\begin{equation}\label{eq:kac_rice}
     P_{\mathbf{w}}=\intop_{-\infty}^{+\infty}\mathrm{d}\mathbf{x}\;\lvert\mathrm{det}\,\mathbf{J}_{\mathbf{w}}\left(\mathbf{x}\right)\rvert\delta\left(\mathbf{v}_{\mathbf{w}}\left(\mathbf{x}\right)\right),
\end{equation}
\noindent where $\mathbf{J}_{\mathbf{w}}\left(\mathbf{x}\right)$ is the Jacobian matrix of $\mathbf{v}_{\mathbf{w}}\left(\mathbf{x}\right)$, evaluated at $\mathbf{x}$. The subscript denotes dependence on the sample ${\mathbf{w}}$. $P_{\mathbf{w}}$ fluctuates from sample to sample even for $N\to\infty$. Roughly speaking, this is because $P_{\mathbf{w}}$ is a product, rather than a sum, of random variables. Accordingly, we expect $P_{\mathbf{w}}\sim\exp\left(N\Sigma^*_q\right)$ for $N$ large, where
\begin{equation}\label{eq:quench_complexity}
    \Sigma^*_q=\lim_{N\to\infty}\frac{1}{N}\mathbb{E}\left[\log P_{\mathbf{w}}\right].
\end{equation}
\noindent The expectation is over $\mathbf{w}$. Thus, the {\em quenched} complexity, $\Sigma^*_q$, describes the asymptotic behavior of the typical value of $P_{\mathbf{w}}$. The {\em annealed} complexity, $\Sigma^*_a$, instead describes the asymptotic behavior of $\mathbb{E}\left[P_{\mathbf{w}}\right]$, i.e., $\mathbb{E}\left[P_{\mathbf{w}}\right]\sim\exp\left(N\Sigma^*_a\right)$. It is always $\Sigma^*_a\geq\Sigma^*_q$.

We evaluate $\Sigma^*_q$ using the replica method \cite{si_link}. The method relies on the computation of $\mathbb{E}\left[P_{\mathbf{w}}^n\right]$ for integer $n$, from which $\mathbb{E}\left[\log P_{\mathbf{w}}\right]=\lim_{n\to0}\;n^{-1}\log\mathbb{E}\left[P_{\mathbf{w}}^n\right]$ (for $n$ real). The major technical difficulty in computing $\mathbb{E}\left[P_{\mathbf{w}}^n\right]$ is to average the replicated determinant of $\mathbf{J}_{\mathbf{w}}\left(\mathbf{x}\right)$ \cite{ben2021counting,ros2023generalized}. That is because $\lvert\mathrm{det}\,\mathbf{J}_{\mathbf{w}}\left(\mathbf{x}\right)\rvert$ is a highly non-linear function of the $w_{ij}$'s.

We note that, when $\mathbf{x}$ is independent of $\mathbf{w}$, the distribution of the eigenvalues of $\mathbf{J}_{\mathbf{w}}\left(\mathbf{x}\right)$ is self-averaging \cite{ahmadian2015properties}. It depends only on $\sigma_w$ and $\frac{1}{N}\sum_i\left(\phi_i'\right)^2$, where $\phi'_i$ is short-hand for the derivative of $\phi\left(x\right)$ evaluated at $x_i$. In our case, $\phi'_i=\Theta\left(x_i\right)$. Hence, $\frac{1}{N}\sum_i\left(\phi_i'\right)^2\equiv f$ is simply the fraction of `active' neurons (i.e., $\phi_i>0$) at $\mathbf{x}$. Then, $\mathbf{J}_{\mathbf{w}}\left(\mathbf{x}\right)$ has one real eigenvalue $\simeq-\sqrt{N}$, $\left(1-f\right)N$ eigenvalues equal to $-1$, and the density of the remaining $fN$ eigenvalues, which henceforth we denote $\lambda_k$ with $k=1,\ldots,fN$, is the uniform distribution within the disc of radius $R=\sigma_w\sqrt{f}$ centered at $-1+\mathrm{i}0$, which henceforth we denote $\rho(z;f,\sigma_w)$. 

The equilibria $\mathbf{x}$ at which one needs to evaluate $\lvert\mathrm{det}\,\mathbf{J}_{\mathbf{w}}\left(\mathbf{x}\right)\rvert$ in Eq. \eqref{eq:kac_rice}, however, {\em depend} on $\mathbf{w}$. To make progress, we assume that this dependence is sufficiently weak so that the density of the $\lambda_k$'s is given by $\rho(z;f,\sigma_w)$ also when $\mathbf{x}$ is an equilibrium of Eq. \eqref{eq:dyn_x} \cite{ahmadian2015properties}. Then, to leading order,
\begin{equation}\label{eq:det_rewrite}    \lvert\mathrm{det}\,\mathbf{J}_{\mathbf{w}}\left(\mathbf{x}\right)\rvert\sim\exp{\left[N\zeta\left(f,\sigma_w\right)\right]},
\end{equation}

\noindent where $\zeta\left(f,\sigma_w\right)\equiv f\int \mathrm{d}z\, \rho(z;f,\sigma_w)\log\lvert z\rvert$. Using Eq.~\eqref{eq:det_rewrite} in Eq.~\eqref{eq:kac_rice}, we obtain

\begin{align}\label{eq:kac_rice_f}
 P_{\mathbf{w}}\sim\intop_{0}^{1} \mathrm{d}f\; & \exp{\left[N\zeta\left(f,\sigma_w\right)\right]}\;\times \\ &\times \intop_{-\infty}^{+\infty}\mathrm{d}\mathbf{x}\;\delta\left(\sum_{i=1}^N\Theta\left(x_i\right)-fN\right)\delta\left(\mathbf{v}_{\mathbf{w}}\left(\mathbf{x}\right)\right)\nonumber,
\end{align}

\noindent where the additional delta function enforces the constraint that the equilibria have a fraction $f$ of active neurons. 
The only term that depends on $\mathbf{w}$ in Eq. \eqref{eq:kac_rice_f} now is $\mathbf{v}_{\mathbf{w}}\left(\mathbf{x}\right)$, and this dependence is linear in the $w_{ij}$'s. 

\begin{figure}[tb]
    \centering
    \includegraphics[scale=0.95]{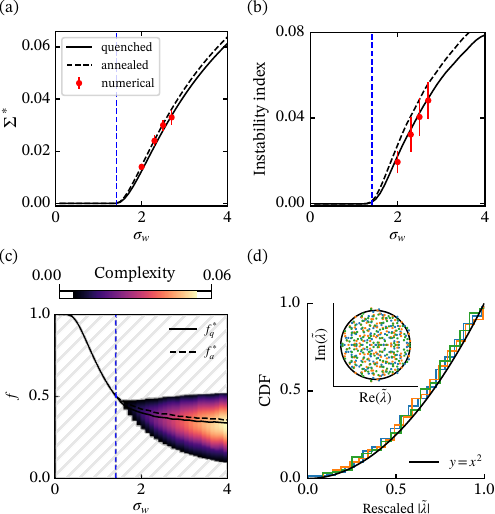}
    \caption{
    (a) $\Sigma^*_q$ (full) and $\Sigma^*_a$ (dashed) as a function of $\sigma_w$. Dots: estimated slope from linear regression ($\log P_N$ vs. $N$); errorbars: SD of the slope estimate ($n=50$ sample networks for each pair of $\sigma_w$ and $N$). 
    (b) Instability index averaged over all the equilibria as a function of $\sigma_w$, according to the quenched (full) and annealed (dashed) theory. Dots and errorbars: mean and SD across networks. 
    (c) $\Sigma_q$ as a function of $\sigma_w$ and $f$. The hatched area denotes $\Sigma_q<0$, the full line the location of $\Sigma^*_q$ and the dashed one that of $\Sigma^*_a$. Vertical dashed line as in (a). 
    (d) Spectral distribution of $\tilde{\lambda}\equiv\left(\sigma_w\sqrt{f}\right)^{-1}\left(1+\lambda\right)$, where $\lambda$ is a non-trivial eigenvalue of $\mathbf{J}_{\mathbf{w}}\left(\mathbf{x}^a\right)$ and $\mathbf{x}^a$ is an equilibrium with a fraction $f^a=f$ of active neurons. Main panel: cumulative distribution function of $\lvert\tilde{\lambda}\rvert$; Inset: full spectrum. Black: theory; Colors: $3$ equilibria randomly chosen from $3$ different sample networks. 
    }
    \label{fig1}
\end{figure}
 Using Eq.~\eqref{eq:kac_rice_f}, $\mathbb{E}\left[P_{\mathbf{w}}^n\right]$ can be evaluated by standard techniques \cite{si_link}. This leads to the introduction of some auxiliary functions of $f$ \cite{si_link}, in particular the {\em overlaps}, $q^{ab}$ $\left(a,b=1,\ldots,n\right)$, defined as

\begin{equation}\label{eq:def_overlaps}
    q^{ab}\left(f\right)\equiv\lim_{N\to\infty}\frac{1}{N}\left(\boldsymbol{\phi}^a\cdot\boldsymbol{\phi}^b\right)=\sqrt{q^{aa}\left(f\right)q^{bb}\left(f\right)}\cos\theta^{ab}\left(f\right),
\end{equation}

\noindent where $\boldsymbol{\phi}^a\equiv\left(\phi\left(x^a_i\right)\right)$, $(\cdot)$ denotes the scalar product, and $\mathbf{x}^a$ and $\mathbf{x}^b$ are two equilibria with the {\em same} $f$. The equality in Eq. \eqref{eq:def_overlaps} follows from  $\lVert\boldsymbol{\phi}^a\rVert\equiv\sqrt{\boldsymbol{\phi}^a\cdot\boldsymbol{\phi}^a}=\sqrt{Nq^{aa}\left(f\right)}$ $(N\to\infty)$; hence, $\cos\theta^{ab}\left(f\right)$ is the cosine similarity (CS) between $\boldsymbol{\phi}^a$ and $\boldsymbol{\phi}^b$.

The $q^{ab}$'s, and the other auxiliary functions, are determined self-consistently by making an explicit assumption about their symmetry under permutations of the indexes $ab$ (replica symmetry) \cite{mezard2009information}. We make the simplest assumption, the replica symmetric ansatz, for {\em all} values of $f$, i.e., $q^{ab}\left(f\right)=Q\left(f\right)\delta_{ab}+q\left(f\right)\left(1-\delta_{ab}\right)$, where $\delta_{ab}$ is the Kronecker symbol. With this, we finally arrive at 

\begin{align}\label{eq:kac_rice_f_SN}
\mathbb{E}\left[P_{\mathbf{w}}^n\right]\sim\intop_{0}^{1}\mathrm{d}f\;\exp{\left[nN\Sigma_q\left(f\right)\right]}\sim\exp{\left[nN\Sigma_q\left(f^*_q\right)\right]},
\end{align}

\noindent for $n\geq2$; $f^*_q$ is the value of $f$ that maximizes $\Sigma_q$. Thus, $\Sigma^*_q=\Sigma_q\left(f^*_q\right)$ (see Eq.~\eqref{eq:quench_complexity}). For $n=1$, we obtain $\mathbb{E}\left[P_{\mathbf{w}}\right]\sim \exp\left[N\Sigma_a\left(f^*_a\right)\right]$ and $f^*_a$ is the value of $f$ that maximizes $\Sigma_a$. We provide explicit expressions for $\Sigma_q\left(f\right)$ and $\Sigma_a\left(f\right)$ in \cite{si_link}.

In Fig.~\ref{fig1}(a), we plot $\Sigma^*_q$ as a function of $\sigma_w$. For $\sigma_w\leq\sqrt{2}$, $\Sigma^*_q=0$ and, therefore, there is a unique equilibrium. Using $\rho(z;f,\sigma_w)$, we compute its instability index, $\alpha\left(f^*_q\right)$ \cite{si_link}. The instability index of an equilibrium $\mathbf{x}^a$ is the fraction of eigenvalues of $\mathbf{J}_{\mathbf{w}}\left(\mathbf{x}^a\right)$ with positive real part. Thus, $\alpha\left(f^*_q\right)N$ is the dimension of the unstable manifold of the equilibrium \cite{eckmann1985ergodic}. We find $\alpha\left(f^*_q\right)=0$ (see Fig.~\ref{fig1}(b)) and, therefore, the equilibrium is stable. Thus, there is a unique, stable equilibrium for $\sigma_w\leq\sqrt{2}$ ($N\to\infty$). 

For $\sigma_w>\sqrt{2}$, $\Sigma^*_q>0$ and increases with $\sigma_w$; hence, there is typically an exponentially large (in $N$) number of equilibria. These equilibria have $0<\alpha\left(f^*_q\right)\ll1$, indicating that they are saddles ($\alpha\left(f^*_q\right)>0$), with a number of unstable directions substantially smaller than $N$ ($\alpha\left(f^*_q\right)\ll1$). For instance, at $\sigma_w=2.5$, less than $4\%$ of the eigenvalues have positive real parts (see Fig.~\ref{fig1}(b)). There exists also an exponential number of equilibria with $f\neq f^*_q$. In fact, $\Sigma_q(f)>0$ over a range of $f$ values, and this range also increases with $\sigma_w$ (Fig.~\ref{fig1}(c)). However, these equilibria are all saddles; there are no stable equilibria for $\sigma_w>\sqrt{2}$ ($N\to\infty$). The network dynamics is chaotic for $\sigma_w>\sqrt{2}$ (see below). Thus, in our model, the transition to chaos is also accompanied by the appearance of exponentially many saddles \cite{wainrib2013topological,ros2023generalized,fournier2025non}.

We compare $\Sigma^*_a$ with $\Sigma^*_q$. For $\sigma_w\leq\sqrt{2}$, $\Sigma^*_a=\Sigma^*_q$, and the maximum occurs for the same value of $f$, i.e., $f^*_a=f^*_q$, while for $\sigma_w>\sqrt{2}$, $\Sigma^*_a>\Sigma^*_q$ (Fig.~\ref{fig1}(a)). In this case, the corresponding maxima occur for $f^*_a>f^*_q$ (Fig.~\ref{fig1}(c)). This indicates the presence of exponentially large, but exponentially rare, fluctuations in $P_{\mathbf{w}}$ across samples. Thus, $\Sigma^*_a$ is not an accurate estimate of the typical number of equilibria. Interestingly, stable equilibria never appear as a result of these large fluctuations. Indeed, $\alpha\left(f\right)>0$ whenever $\Sigma_a\left(f\right)>0$ \cite{si_link}. We note that the difference between $\Sigma^*_a$ and $\Sigma^*_q$ is quantitatively small in the range of $\sigma_w$ that we have explored. 

We seek to validate our theory numerically by searching the equilibria of sample networks with a root-finding algorithm (Newton's method with backtracking). As we explain in \cite{si_link}, in our case (i.e., $\phi\left(x\right)=x\Theta\left(x\right)$) we can unequivocally decide whether the `equilibrium' returned by the algorithm is a {\em true} equilibrium of Eq.~\eqref{eq:dyn_x}. Extensive search (as described in \cite{si_link}) is feasible only for relatively small networks ($N\leq200$).

We start by computing estimates of $\Sigma^*_q$ as a function of $\sigma_w$ from the equilibria found numerically.  The resulting estimates are plotted in Fig.~\ref{fig1}(a) on top of the theoretical prediction. There is a fairly good agreement with the theory, despite the small values of $N$. In Fig.~\ref{fig1}(d), we compare instead the spectral distribution of $\mathbf{J}_{\mathbf{w}}\left(\mathbf{x}^a\right)$, where $\mathbf{x}^a$ is an equilibrium, with the theoretical prediction.  Again, there is a very good agreement. We see this also in Fig.~\ref{fig1}(b), where we compare the instability index, averaged over all the equilibria $\mathbf{x}^a$, computed by numerically diagonalizing  $\mathbf{J}_{\mathbf{w}}\left(\mathbf{x}^a\right)$, with the average instability index computed from the theory, which is simply $\alpha\left(f^*_q\right)$ for $N\to\infty$. 

\begin{figure}[tb]
    \centering
    \includegraphics[scale=0.95]{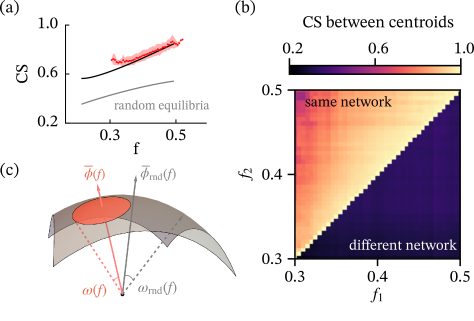}
    \caption{\small 
    (a) Cosine similarity (CS) between network's equilibria and their centroid as a function of $f$. Full line: theory; Red dots: estimates from the equilibria numerically found (shaded area is $\pm$SD across networks); Gray line: CS between random equilibria and their centroid. Parameters: $\sigma_w=2.5$ and $N=200$. 
    (b) CS between centroids at different $f_1$ and $f_2$, both in the same network (above diagonal), or in different networks (below diagonal). Parameters as in (a).
    (c) Cartoon illustrating the spatial organization of actual and random equilibria. 
    }
    \label{fig2}
\end{figure}

The overlaps $q^{ab}(f)$ encode information about the spatial organization of the equilibria. Let us consider the equilibria $\mathbf{x}^a$ with the same $f^a=f$. The corresponding $\boldsymbol{\phi}^a$'s lay inside a spherical cap \footnote{For $N\to\infty$, the volume of the cap, and hence the equilibria, will be concentrated in the rim.} on the $N-$sphere of radius $\sqrt{NQ\left(f\right)}$ centered at the origin (in the non-negative orthant); see Eq.~\eqref{eq:def_overlaps}. The pole of this spherical cap is given by the intersection of the $N-$sphere and the straight line defined by the `centroid', $\overline{\boldsymbol{\phi}}\left(f\right)$, that is, the average of the $\boldsymbol{\phi}^a$'s. The polar angle, $\omega\left(f\right)$, satisfies $\cos\omega\left(f\right)=\sqrt{q\left(f\right)/Q\left(f\right)}$. This is illustrated in Fig.~\ref{fig2}(a), where we plot, as a function of $f$, the CS between the equilibria numerically found and their centroid (red dots) together with the theoretical prediction, i.e., $\cos\omega\left(f\right)$ (black line) for $\sigma_w=2.5$. As can be seen, there is good agreement.

This geometrical organization is natural in high-dimensional spaces ($N\to\infty$). To see this, consider `random' equilibria, i.e., random vectors whose components are independently and identically distributed such that: (i) components are positive with probability $f$ and $0$ otherwise; (ii) their average is $1$, as the average $\phi^a_i$ over equilibria and neurons; (iii) their variance is $Q\left(f\right)-1$. It is easy to verify that also the random equilibria with a given $f$ lay inside a spherical cap on the $N-$sphere of radius $\sqrt{NQ\left(f\right)}$, in the non-negative orthant \cite{si_link}. The centroid, defining the pole, is the vector with all components equal to $1$; the polar angle, $\omega_\mathrm{rnd}\left(f\right)$, satisfies $\cos\omega_\mathrm{rnd}\left(f\right)=\sqrt{1/Q\left(f\right)}$ (gray line in Fig.~\ref{fig2}(a); for $\sigma_w=2.5$).

In the whole range of $\sigma_w>\sqrt{2}$ that we have explored, it is always $q\left(f\right)>1$ for all $f$ such that $\Sigma_q\left(f\right)>0$ and, therefore, $\omega_\mathrm{rnd}\left(f\right)>\omega\left(f\right)$; compare the black and gray lines in Fig.~\ref{fig2}(a). Hence, for any $f$, the cap containing the equilibria is smaller than the cap containing the corresponding random equilibria (see Fig.~\ref{fig2}(c)). This is the geometrical manifestation of the correlations between equilibria. These originate from the fact that $\boldsymbol{\phi}^a$'s are all equilibria of the {\em same} sample network. In fact, $x^a_i$ and $x^b_i$, and hence $\phi^a_i$ and $\phi^b_i$, (i.e., same neuron, different equilibria), are correlated because of the {\em quenched} couplings $w_{ij}$'s \footnote{Correlations between $x^a_i$ and $x^a_j$ (i.e., different neurons, same equilibrium) are negligible because of the mean-field nature of the model.}. 

According to the above argument, we expect correlations also between equilibria with different values of $f$, i.e., $f_1$ and $f_2$. A simple way to quantify these correlations is to compute the CS between the corresponding centroids, $\overline{\boldsymbol{\phi}}\left(f_1\right)$ and $\overline{\boldsymbol{\phi}}\left(f_2\right)$. In fact, this CS encodes the average overlap, $q\left(f_1,f_2\right)$, between equilibria $\boldsymbol{\phi}^a$ with $f^a=f_1$ and equilibria $\boldsymbol{\phi}^b$ with $f^b=f_2$, i.e., 
\begin{equation}\label{eq:cossim_equilibria}
    \cos\overline{\omega}\left(f_1,f_2\right)=\frac{q\left(f_1,f_2\right)}{\sqrt{q\left(f_1\right)q\left(f_2\right)}}.
\end{equation}

\noindent For random equilibria, $q\left(f_1,f_2\right)=1$. We have estimated the CS between $\overline{\boldsymbol{\phi}}\left(f_1\right)$ and $\overline{\boldsymbol{\phi}}\left(f_2\right)$ using the equilibria found numerically; the results are illustrated in Fig.~\ref{fig2}(b). As can be seen, the estimated CSs are substantially larger than those expected if equilibria at $f_1$ and $f_2$ were random; equilibria are strongly correlated for each $(f_1,f_2)$. Geometrically, this implies that equilibria `occupy' in the phase space a much smaller volume than random ones \footnote{Clearly, equilibria, random or otherwise, fill up zero volume. What we mean is that the actual equilibria can be enclosed in a volume that is substantially smaller than the volume needed to enclose the corresponding random equilibria}. We note that the strong correlations are not inherited from correlations in the synaptic connectivity, in any obvious way at least, since the synaptic connectivity is completely random.

We study the network dynamics using DMFT \cite{si_link}. In steady states, DMFT determines self-consistently the autocorrelation function
\begin{equation}
    C\left(\tau\right)\equiv\lim_{N\to\infty}\frac{1}{N}\left\langle\boldsymbol{\tilde{\phi}}^t\cdot\boldsymbol{\tilde{\phi}}^{t+\tau}\right\rangle
\end{equation}

    \noindent with $\boldsymbol{\tilde{\phi}}^t\equiv\left(\phi\left(\tilde{x}_i\left(t\right)\right)\right)$ where the $\mathbf{\tilde{x}}\left(t\right)$'s are solutions of Eq.~\eqref{eq:dyn_x} (compare with Eq.~\eqref{eq:def_overlaps}), and $\langle\cdot\rangle$ denotes the long-time average or, equivalently, the average over the invariant (physical) measure. For $N\to\infty$, $C\left(\tau\right)$ only depends on $\sigma_w$.

For $\sigma_w\leq\sqrt{2}$, $C\left(\tau\right)=Q_\mathrm{S}$ is the only solution of DMFT. It corresponds to a unique, stable equilibrium and $Q_\mathrm{S}=Q\left(f^*_q\right)$. For $\sigma_w>\sqrt{2}$, this solution still exists. However, there is only one $Q_\mathrm{S}$, as before, and it is different from $Q\left(f^*_q\right)$ now. Thus, DMFT does not reveal all the equilibria of Eq.~\eqref{eq:dyn_x} (i.e., the ones with $Q\neq Q_\mathrm{S}$) and those revealed are {\em not} the typical ones. For $\sigma_w>\sqrt{2}$, there are additional solutions which do not correspond to equilibria, i.e., $C\left(\tau\right)$ is not constant. We only consider the stable solution with $\lim_{\tau\to\infty}C\left(\tau\right)=q_\mathrm{D}$, which describes the chaotic dynamics of Eq.~\eqref{eq:dyn_x} for $\sigma_w>\sqrt{2}$ (starting from random initial conditions). 

\begin{figure}[tb]
    \centering
    \includegraphics[scale=1.0]{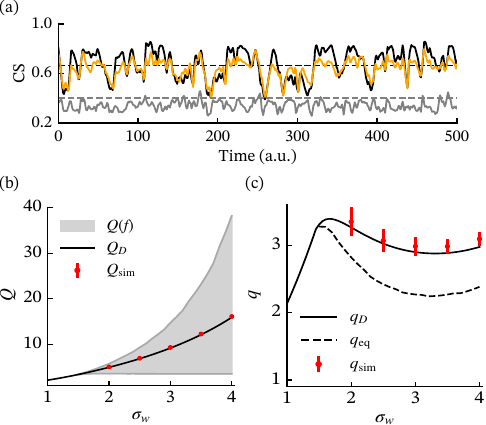}
    \caption{(a) Cosine similarity between $\boldsymbol{\tilde{\phi}}^t$ and $\langle\boldsymbol{\tilde{\phi}}\rangle$ (black) and between $\boldsymbol{\tilde{\phi}}^t$ and $\boldsymbol{\bar{\phi}}$ (orange), when all vectors are computed in the {\em same} sample network ($\sigma_w=2.5$, $N=200$). For comparison, we also plot the cosine similarity between $\boldsymbol{\tilde{\phi}}^t$ and $\boldsymbol{\bar{\phi}}$, when these vectors are computed in {\em different} sample networks (gray).   
    (b) $Q_\mathrm{D}$ as a function of $\sigma_w$. Red dots with errorbars (SD) represent numerical estimates ($n=50$ samples with $N=5000$). The gray area represents the range of $Q(f)$ over the corresponding equilibria. 
    (c) $q_\mathrm{D}$ (full) and $q_\mathrm{eq}$ (dashed) as a function of $\sigma_w$. Red dots as in (b).}                                      
    \label{fig3}
\end{figure}

We use the geometrical construction described above for the equilibria to `visualize' the chaotic dynamics. To that end, we find it convenient to define $Q_\mathrm{D}\equiv C\left(0\right)$. Clearly, $Q_\mathrm{D}>q_\mathrm{D}$. Then, the $\boldsymbol{\tilde{\phi}}^t$'s lay inside a cap on the $N-$sphere of radius $\sqrt{NQ_\mathrm{D}}$ centered at the origin, again in the non-negative orthant. The pole of the cap is given by the intersection of this $N-$sphere and the straight line defined by the centroid $\langle\boldsymbol{\tilde{\phi}}\rangle$, i.e., the average over time of $\boldsymbol{\tilde{\phi}}^t$. The polar angle satisfies $\cos\omega_\mathrm{D}=\sqrt{q_\mathrm{D}/Q_\mathrm{D}}$. This is illustrated in Fig.~\ref{fig3}(a), where we plot the time course of the CS between $\boldsymbol{\tilde{\phi}}^t$ and $\langle\boldsymbol{\tilde{\phi}}\rangle$.

To locate the `dynamical' cap (i.e., the cap that contains the attractor) with respect to the equilibria, we also plot in Fig.~\ref{fig3}(a) the CS between $\boldsymbol{\tilde{\phi}}^t$ and $\overline{\boldsymbol{\phi}}$, defined as the average of {\em all} the equilibria in the network (note that  $\overline{\boldsymbol{\phi}}\to\overline{\boldsymbol{\phi}}\left(f^*_q\right)$ for $N\to\infty$). As can be seen, the two CSs are highly correlated, indicating that $\langle\boldsymbol{\tilde{\phi}}\rangle$ and $\overline{\boldsymbol{\phi}}$ are (almost) collinear. Does the dynamical cap lay inside the region that contains the equilibria?

To answer this question, we begin by determining whether there exists an $f_\mathrm{eq}$ such that $Q\left(f_\mathrm{eq}\right)=Q_\mathrm{D}$. Clearly, if such an $f_\mathrm{eq}$ does not exist, then the dynamical cap is outside. It turns out that a unique $f_\mathrm{eq}$ always exists. This is illustrated in Fig.~\ref{fig3}(b), where we plot $Q_\mathrm{D}$ as a function of $\sigma_w$, together with the largest, $Q_\mathrm{max}$, and the smallest, $Q_\mathrm{min}$, value of $Q$ over the equilibria with the same $\sigma_w$. As can be seen, $Q_\mathrm{min}<Q_\mathrm{D}<Q_\mathrm{max}$ in the whole range of $\sigma_w>\sqrt{2}$ that we have explored. There is a unique $f_\mathrm{eq}$ because $Q\left(f\right)$ is a strictly monotonous decreasing function of $f$ \cite{si_link}. We note that it is always $Q_\mathrm{D}>Q\left(f^*_q\right)$, and hence $f_\mathrm{eq}<f^*_q$, for $\sigma_w>\sqrt{2}$ (not shown). 

Next, using $f_\mathrm{eq}$, we determine $q_\mathrm{eq}\equiv q\left(f_\mathrm{eq}\right)$ to be compared with $q_\mathrm{D}$. This is illustrated in Fig.~\ref{fig3}(c), where we plot both $q_\mathrm{D}$ and $q_\mathrm{eq}$ as a function of $\sigma_w$. As can be seen, in the whole range of $\sigma_w>\sqrt{2}$ explored, it is $q_\mathrm{D}>q_\mathrm{eq}$. The dynamical cap lays always inside the region containing the equilibria, at a distance from the origin determined by $Q_\mathrm{D}$. As already noticed for other models \cite{stubenrauch2025fixed,fournier2025non}, also in our case the dynamics is outside the cap containing the typical equilibria, because $Q_\mathrm{D}>Q\left(f^*_q\right)$. 

\begin{figure}[tb]
    \centering
    \includegraphics[scale=1.0]{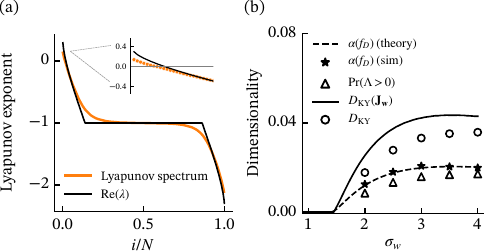}
    \caption{
    (a) Lyapunov spectrum of the dynamics for $\sigma_w=2.5$ and $N=500$ (orange) compared with the real parts of $\mathbf{J}_{\mathbf{w}}(\tilde{\mathbf{x}})$ eigenspectrum (black) computed using random-matrix theory. Inset: largest 5\% of exponents/eigenvalues. 
    (b) Dimension of the attractor as a function of $\sigma_w$. Symbols: numerical estimates ($N=2000$); Lines: theory. See main text for details.}                                      
    \label{fig4}
\end{figure}

We numerically estimated the Lyapunov exponents (LEs) for different values of $\sigma_w>\sqrt{2}$ using Benettin's algorithm \cite{benettinetal1980b,si_link} (see also \cite{engelken2023lyapunov}). The result for $\sigma_w=2.5$ is shown in Fig.~\ref{fig4}(a). As seen more clearly in the inset, a fraction of LEs are positive; this fraction is independent of $N$ (see Fig.~S4(c)~\cite{si_link}). This indicates sensitive dependence on the initial conditions due to the presence of an {\em extensive} number of globally expanding degrees of freedom.  

In the same figure, we also plot the real parts of $\mathbf{J}_{\mathbf{w}}\left(\mathbf{\tilde{x}}\right)$ spectrum computed using random matrix theory \cite{ahmadian2015properties}, i.e., the non-trivial part of the spectrum is described by $\rho(z;f_\mathrm{D},\sigma_w)$, where $f_\mathrm{D}$ is the fraction of active neurons on the attractor. Indeed, the spectrum is self-averaging on the attractor. This is because, for $N\to\infty$, all points, $\mathbf{\tilde{x}}$, on the attractor have the same fraction of active neurons, $f=f_\mathrm{D}$, and the correlations between $\mathbf{w}$ and $\mathbf{\tilde{x}}$ can be neglected (see Fig.~S4(a)~\cite{si_link}).

Notably, the real parts of $\mathbf{J}_{\mathbf{w}}\left(\mathbf{\tilde{x}}\right)$ spectrum provide a fairly accurate approximation to the LEs. A theoretical rationale for this correspondence is provided in \cite{si_link}. Here, we exploit this observation; indeed, Fig.~\ref{fig4}(a) suggests that one can obtain useful approximations to various quantities defined via the LEs -- which are notoriously difficult to compute -- by utilizing instead the real parts of the Jacobian spectrum, far more tractable. This expectation is confirmed in Fig.~\ref{fig4}(b) for different values of $\sigma_w>\sqrt{2}$.

We computed the fraction of positive LEs -- the fraction of globally expanding degrees of freedom -- and the Kaplan-Yorke attractor dimension, $D_\mathrm{KY}$ -- the fractional dimension of the subspace where the dynamics is volume-preserving \cite{eckmann1985ergodic}. Both quantities are positive and independent of $N$ (see Fig.~S4(c)~\cite{si_link}), confirming that the dynamics exhibits extensive spatio-temporal chaos for $\sigma_w>\sqrt{2}$. Consistent with related models \cite{engelken2023lyapunov}, the attractor dimension is substantially smaller than the dimension of the phase space ($D_\mathrm{KY}<0.04$ for $\sigma_w\leq4$). Next, we computed the same quantities by using the real parts of $\mathbf{J}_{\mathbf{w}}\left(\mathbf{\tilde{x}}\right)$ spectrum. As can be seen in the figure, these proxies yield fairly tight upper bounds across the whole range of $\sigma_w>\sqrt{2}$ explored.

In summary, we have shown that the attractor is confined to a relatively small region in the phase space, determined by the equilibria. Because of this geometric organization, the quantitative features of these equilibria provide natural bounds on network's dynamics. Specifically, the overlap between points on the attractor, $q_\mathrm{D}$, is larger than the overlap between equilibria at the same distance from the origin, $q_\mathrm{eq}$ (Fig.~\ref{fig3}(c)). Consequently, $q_\mathrm{eq}$ is a lower bound on the level of spatial heterogeneity (i.e., variability in the $\langle\tilde{\phi}_i\rangle$) generated by the dynamics, measured by $q_\mathrm{D}$. Similarly, the fraction of positive Lyapunov exponents is bounded from above by the fractional dimension of the unstable manifold of points on the attractor, $\alpha\left(f_\mathrm{D}\right)$ (Fig.~\ref{fig4}(b)). In turn, $\alpha\left(f_\mathrm{D}\right)\leq\alpha\left(f^*_q\right)$, with equality holding at the onset of chaos. The attractor dimension is therefore fundamentally constrained by the instability index of the typical equilibria.

In our model, even the equilibrium with the highest number of unstable directions exhibits an instability index substantially smaller than $1$ (see \cite{si_link}, Fig.~S3). Intriguingly, such small instability indices appear to be typical for a large class of mean-field models \cite{ben2021counting}. This explains why the dynamics in these models \cite{engelken2023lyapunov,clark2023dimension,stubenrauch2025fixed}  can be described by a {\em fractionally} small number of effective degrees of freedom\footnote{The attractor dimension, however, remains extensive. That is, the {\em number} of effective degrees of freedom needed to describe the dynamics is proportional to $N$. Hence, for any practical purpose such as, e.g., making predictions using purely observational data, the dynamics cannot be considered 'low-dimensional' \cite{cecconi2012predicting}.}.

Our approach can be extended to more `structured' networks, such as partially symmetric networks \cite{marti2018correlations,Berlemont2022} and Hopfield-like memory networks \cite{treves1988metastable}. These architectures provide core models for experimental systems in Neuroscience. In such networks, the dynamics are more complex, DMFT cannot be solved analytically, and ergodicity can be broken. A fundamental question is therefore: can we predict the breaking of ergodicity from the geometric organization of the equilibria? 

This work was partially supported by NIH/NINDS award 1UF1NS115779 (GLC). We would like to thank Tatiana Engel, David Hansel, Alessandro Ingrosso, Misha Katkov, Misha Tsodyks and Pierfrancesco Urbani for useful discussions. 
    
\bibliography{refs.bib}
\end{document}


\onehalfspacing       
\title{Supplemental Material for ``On the relationship between equilibria and dynamics in large, random neuronal networks''}

\author{Xiaoyu Yang}
\affiliation{Graduate Program in Physics and Astronomy, Stony Brook University, Stony Brook, NY, USA}
\affiliation{Department of Neurobiology and Behavior, Stony Brook University, Stony Brook, NY, USA}
\affiliation{Center for Neural Circuit Dynamics, Stony Brook University, Stony Brook, NY, USA}

\author{Giancarlo La Camera}
\email{giancarlo.lacamera@stonybrook.edu}
\affiliation{Department of Neurobiology and Behavior, Stony Brook University, Stony Brook, NY, USA}
\affiliation{Center for Neural Circuit Dynamics, Stony Brook University, Stony Brook, NY, USA}
\affiliation{AI Innovation Institute, Stony Brook University, Stony Brook, NY, USA}
\affiliation{Center for Advanced Computational Science, Stony Brook University, Stony Brook, NY, USA}
\affiliation{Graduate Programs in Neuroscience, Stony Brook University, Stony Brook, NY, USA}

\author{Gianluigi Mongillo}

\email{gianluigi.mongillo@gmail.com}
\affiliation{Sorbonne Universit\'e, INSERM, CNRS, Institut de la Vision, F-75012 Paris, France}
\affiliation{Centre National de la Recherche Scientifique, Paris, France}
\affiliation{School of Natural Sciences, Institute for Advanced Study, Princeton, NJ, USA.}

\maketitle
\tableofcontents
\clearpage            

\section{The model}

In this section, we describe the model neuronal network that we will be studying and briefly discuss its relevance to Neuroscience. The network consists of $N$ all-to-all connected neurons and is meant to describe a local cortical network. The instantaneous state of neuron $i=1,\ldots,N$ is described by $x_i$, which evolves in time according to
\begin{equation}
    \label{eq: network_dynamics}
    \dot{x}_i = -x_i + \sqrt{N} + \frac{1}{\sqrt{N}}\sum_{j=1}^Nw_{ij}\phi(x_j).
\end{equation}

\noindent where $w_{ij}$ is the efficacy of the synapse connecting neuron $j$ to neuron $i$ and $\phi\left(x\right)$ is the so-called activation function (or $f$-$I$ curve), which determines the neuron's firing rate as a function of its instantaneous state.

We take the $w_{ij}$'s to be independent Gaussian deviates with $\mathbb{E}\left[w_{ij}\right]=-1$ and $\mathrm{Var}\left[w_{ij}\right]=\sigma^2_w$. This is meant to capture, at least in a first approximation, the heterogeneity of the synaptic connectivity observed in cortical networks. 

We take $\phi(x)\equiv x\Theta(x)$, where $\Theta(x)$ is the Heaviside step function: $\Theta(x) =1$ for $x \ge 0$, and $\Theta(x)=0$ for $x<0$. Therefore, our neurons are threshold-linear units; the firing rate is non-negative and does not saturate for large $x_j$. It has been argued that threshold-linear units provide a better description of the activity of cortical neurons in physiological conditions \cite{treves1990graded} as compared to sigmoidal functions such as, e.g., the hyperbolic tangent. Indeed, besides the fact that firing rates are obviously non-negative quantities, neurons in the cortex fire at rates ($\sim 1$-$10$Hz) substantially below saturation ($\sim 100$Hz). Furthermore, in this range of `physiological' rates, a threshold-linear activation function provides a reasonably good fit to the activation function of real neurons \cite{stevens1998input}.   

Eq.~\eqref{eq: network_dynamics} can be interpreted as describing the `leaky' integration of the total synaptic input. In this interpretation, we can think of $x_i$ as the synaptic current (conductance) to (of) neuron $i$ or the membrane potential of neuron $i$. This interpretation can be formally justified in some limiting cases (see, e.g., \cite{harish2015asynchronous}). We simply take Eq.~\eqref{eq: network_dynamics} as a useful phenomenological model that describes neuronal (spiking) activity on some intermediate time scale (i.e., $100$ms-$1$s).

In this interpretation, the synaptic input to neuron $i$ is the sum of two contributions: an external, time-independent {\em excitatory} input (i.e., the term $+\sqrt{N}$ in Eq.~\eqref{eq: network_dynamics}) and the recurrent synaptic input, generated by the activity within the network (i.e., the sum over $j$). This latter is the sum of the firing rates of all the other neurons in the network, weighted by the synaptic weights $w_{ij}$'s. Because the $w_{ij}$'s are negative on average, the input generated by the recurrent connectivity is dominated by {\em inhibition}, as we now discuss in more detail. 

Since $w_{ij} = -1+\sigma_w z_{ij}$, where $z_{ij}$ are independent, standard Gaussian deviates, we can write the total input as 
\begin{equation} \label{eq:balance0}
\sqrt{N} + \frac{1}{\sqrt{N}}\sum_{j=1}^Nw_{ij}\underbrace{\phi(x_j)}_{\equiv\phi_j}=\sqrt{N}\left(1 - \frac{1}{N}\sum_{j=1}^N \phi_j\right) + \frac{\sigma_w}{\sqrt{N}}\sum_{j=1}^N z_{ij}\phi_j.
\end{equation}

As we show in detail in the following, the steady states of the network dynamics (i.e., equilibria and chaotic dynamics) are such that both these terms remain $\mathcal{O}\left(1\right)$, and hence comparable in size, in the limit $N\to\infty$. For this to be possible, it must be   
\begin{equation} \label{eq:mu}
\lim_{N\to\infty}\;\sqrt{N} \left (1 - \frac{1}{N}\sum_{j=1}^N\langle\phi_j\rangle\right)\equiv\mu=\mathcal{O}\left(1\right) 
\end{equation}

\noindent where $\langle\cdot\rangle$ denotes long-time averaging. Clearly, in an equilibrium $\langle\phi_j\rangle=\phi_j$. Eq.~\eqref{eq:mu} indicates that, in large $N$ limit, the average firing rate (over neurons and over time) in the network dynamically adjusts such that the excitatory and inhibitory component of the synaptic inputs to each neuron cancel to leading order, and the {\em net} input, $\mu$, remains $\mathcal{O}\left(1\right)$. For $N$ suitably large, Eq.~\eqref{eq:mu} can be equivalently rewritten as
\begin{equation} \label{eq:balance2-0}
\frac{1}{N} \sum_{j=1}^N \langle\phi_j\rangle = 1 - \frac{\mu}{\sqrt{N}}.
\end{equation}

\noindent Hence, the `balance condition' Eq.~\eqref{eq:mu} fixes the average firing rate in the network for $N\to\infty$. Note that the average firing rate cannot change across steady states, i.e., it is always $1$ (because of our choice of parameters). Instead, $\mu$ can, and will, change across steady states. 

The balance condition Eq.~\eqref{eq:mu} and its consequences for the network dynamics are exactly the same as in `standard' balanced networks \cite{van1996chaos,renart2010asynchronous}. In `standard' balanced networks, one requires that all the $w_{ij}$'s with the same $j$, i.e., all the synapses originating from neuron $j$, have the same sign, i.e., they are all excitatory or inhibitory. This is to comply with neurophysiological observations, the so-called Dale's principle, which indicate that, typically, a pre-synaptic neuron has the same effect, excitatory or inhibitory, on all its post-synaptic targets. Clearly, this is not true in our model network, because the $w_{ij}$ with the same $j$ can be both positive and negative. However, as we have shown, the sign-constraint can be safely disregarded as long as one is interested in studying the balanced regime (in the limit $N\to\infty$). This also results in slightly simplified models.

\section{Counting equilibria using the replica method} \label{sec:kac}
\subsection{Preliminaries}

 The system described by Eq.~\eqref{eq: network_dynamics} exhibits a {\em paradigmatic transition} between a unique, stable equilibrium and a chaotic attractor at $\sigma_w^{(c)}=\sqrt{2}$ (this result can be obtained via the DMFT presented in Sec.~\ref{sec:dmft}. See also \cite{Berlemont2022}). 
 
For a given realization of the synaptic matrix $\mathbf{w}$, the total number of equilibria of Eq.~\eqref{eq: network_dynamics}, denoted as $P_{\mathbf{w}}$, can be computed using the Kac-Rice formula
\begin{equation}
    \label{eq: kac-rice}
    P_{\mathbf{w}} = \int_{-\infty}^{\infty}\left(\prod_{i=1}^N\mathrm{d}x_i\right)
    \prod_{i=1}^N\delta\Big(-x_i+\sqrt{N}+\frac{1}{\sqrt{N}}\sum_{j=1}^N w_{ij}\phi_j\Big)|\det\mathbf{J}_{\mathbf{w}}(\mathbf{x})|, 
\end{equation}
where $\mathbf{J}_{\mathbf{w}}(\mathbf{x})$ is the Jacobian matrix of the velocity field, $v_i\left(\mathbf{x}\right) \doteq -x_i+\sqrt{N}+\frac{1}{\sqrt{N}}\sum_{j=1}^Nw_{ij}\phi_j$, evaluated at $\mathbf{x}$. $P_{\mathbf{w}}$ is a random variable that fluctuates across realizations of $\mathbf{w}$. However, its logarithm per site, $N^{-1}\log P_{\mathbf{w}}$, is expected to be a {\em self-averaging} quantity in the limit $N\to\infty$. This defines the {\em quenched complexity} 
\begin{equation}
    \Sigma_q^* = \lim_{N\to\infty}\frac{1}{N}\log P_{\mathbf{w}} = \lim_{N\to\infty}\frac{1}{N}\mathbb{E}[\log P_{\mathbf{w}}],
\end{equation}
where $\mathbb{E}[\cdot]$ denotes the average over $w_{ij}$. The quenched complexity $\Sigma_q^*$ describes the asymptotic behavior of the typical value of $P_{\mathbf{w}}$. Alternatively, the asymptotic behavior of $\mathbb{E}[P_{\mathbf{w}}]$ defines the {\em annealed complexity} 
\begin{equation}
    \Sigma_a^* = \lim_{N\to\infty}\frac{1}{N}\log \mathbb{E}[P_{\mathbf{w}}].
\end{equation}

Evaluating $\Sigma_q^*$ requires calculating the average of logarithms, which is typically a challenging task. To compute the quenched complexity, we use the replica method by first evaluating the average of $P^n_{\mathbf{w}}$ for integer $n$ and then taking the limit $n\to 0$. This yields
\begin{equation}
    \Sigma_q^* = \lim_{N\to\infty}\lim_{n\to 0}\frac{1}{Nn}\left(\mathbb{E}[P^n_{\mathbf{w}}]-1\right)
    =\lim_{n\to 0}\lim_{N\to \infty}\frac{1}{Nn}\log\mathbb{E}[P^n_{\mathbf{w}}], 
\end{equation}
where we adopted the common assumption that the two limits, $n\to 0$ and $N\to\infty$, can be exchanged safely. 

Using the Fourier transforms of the $\delta$-functions, the replicated Kac-Rice formula is given by 
\begin{equation}
    P^n_{\mathbf{w}}
    =\int_{-\infty}^{\infty}\left(\prod_{a=1}^n\prod_{i=1}^N\frac{\mathrm{d}x_i^a \mathrm{d}\tilde{x}_i^a}{2\pi}\right)
    \exp\left[\sum_{a=1}^{n}\sum_{i=1}^N i\tilde{x}_i^a(x_i^a-\sqrt{N}-\frac{1}{\sqrt{N}}\sum_{j=1}^N w_{ij}\phi_j^a) \right]
    \times \prod_{a=1}^n |\det \mathbf{J}_{\mathbf{w}}(\mathbf{x}^a)|, 
\end{equation}
We parameterize the synaptic weights as $w_{ij}=-1+\sigma_w z_{ij}$, where $z_{ij}\sim\mathcal{N}(0,1)$ are standard Gaussian variables. We define $\mu^a=\sqrt{N}\big(1-\frac{1}{N}\sum_{i=1}^N\phi_i^a\big)$ as the mean input to replica $a$. After collecting terms with same coefficient $z_{ij}$, we arrive at the following expression 
\begin{equation}
    \label{eq: replicated-kac-rice}
    P^n_{\mathbf{w}}
    =\int_{-\infty}^{\infty}\left(\prod_{a=1}^n\prod_{i=1}^N\frac{\mathrm{d}x_i^a \mathrm{d}\tilde{x}_i^a}{2\pi}\right)\exp\left[\sum_{a=1}^{n}\sum_{i=1}^N i\tilde{x}_i^a(x_i^a-\mu^a) + \sum_{ij} z_{ij}\Big(\frac{\sigma_w}{\sqrt{N}}\sum_{a=1}^n i\tilde{x}_i^a\phi_j^a\Big)\right]
    \prod_{a=1}^n |\det \mathbf{J}_{\mathbf{w}}(\mathbf{x}^a)|. 
\end{equation}
The Jacobian determinants in the last term are evaluated at different states $\mathbf{x}^a$ but for the same realization $\mathbf{w}$, hence, they are correlated with each other and with the delta functions (in the Fourier form). Computing the expectation of such a product is hard and often impractical. A previous study, however, suggested that, in the limit $N\to\infty$, the empirical spectral density of the Jacobian of certain disordered systems is {\em self-averaging} \cite{ahmadian2015properties}. Their theory applies to our model network in Eq.~\eqref{eq: network_dynamics}. It turns out that, for a state $\mathbf{x}$ with $Nf$ active neurons, $\mathbf{J}_{\mathbf{w}}(\mathbf{x})$ has one real eigenvalue of $-\sqrt{N}$, $(1-f)N$ eigenvalues of $-1$, and the remaining $fN$ eigenvalues are uniformly distributed within the disc of radius $R=\sigma_w\sqrt{f}$ centered at $-1+i0$. We denote this nontrivial distribution as $\rho(z;f,\sigma_w)$. 

The absolute value of the Jacobian determinant, $|\det \mathbf{J}(\mathbf{x})|=\prod_{i=1}^N |\lambda_i|$, is governed by the $fN$ nontrivial eigenvalues. For large $N$, we approximate the determinant using the empirical spectral density $\rho(z;f,\sigma_w)$
\begin{equation}
    |\det \mathbf{J}(\mathbf{x})| 
    = \exp\left[Nf\cdot \frac{1}{Nf}\sum_{i=1}^{N} \ln |\lambda_i|\right]
    \overset{N\to\infty}{\approx} \exp\left[Nf\int \mathrm{d}z \rho(z;f,\sigma_w)\log |z|\right]. 
\end{equation}
To leading order, this simplifies to
\begin{equation}
    \label{eq: asymptotic_jacobian}
    |\det \mathbf{J}(\mathbf{x})| \sim \exp\left[N\zeta(f,\sigma_w)\right], 
\end{equation}
where $\zeta(f,\sigma_w)\equiv f\int \mathrm{d}z \rho(z;f,\sigma_w)\log|z|$. We assume that the spectral density $\rho(z;f,\sigma_w)$ is weakly correlated with the delta functions (which impose the equilibrium condition). This allows us to use the asymptotic relation in Eq.~\eqref{eq: asymptotic_jacobian} also when $\mathbf{x}$ is an equilibrium of the system. Using the residue theorem, one can derive the following analytical expression for $\zeta(f,\sigma_w)$:
\begin{equation}
    \zeta(f,\sigma_w) = \begin{cases}
        \frac{1}{2\sigma_w^2} + \frac{f}{2}\ln(\sigma_w^2f)-\frac{f}{2} & \sigma_w\sqrt{f}>1, \\
        0 & \sigma_w\sqrt{f}\le 1. 
    \end{cases}
\end{equation}
Furthermore, we make the assumption that the Jacobian determinants corresponding to different replicas are also weakly correlated for large $N$. This implies that the product of the determinants in Eq.~\eqref{eq: replicated-kac-rice} has a leading order of
\begin{equation}
    \prod_{a=1}^n |\det \mathbf{J}_{\mathbf{w}}(\mathbf{x}^a)| \sim \exp \left[N\sum_{a=1}^n\zeta(f_a,\sigma_w)\right]. 
\end{equation}
We note that now the only term that depends on $\mathbf{w}$ in Eq.~\eqref{eq: replicated-kac-rice} is $\exp\left[\sum_{ij}z_{ij}(\frac{\sigma_w}{\sqrt{N}}\sum_{a=1}^n i\tilde{x}_i^a\phi_j^a)\right]$, which is linear in the $z_{ij}$'s. 

\paragraph{Instability index}
The stability of an equilibrium $\mathbf{x}^a$ is  determined by its corresponding Jacobian matrix $\mathbf{J}(\mathbf{x}^a)$: the equilibrium is stable if all eigenvalues have non-positive real parts and unstable otherwise. We quantify this stability using the {\em instability index} $\alpha$ ($0\le \alpha\le 1$), defined as the fraction of eigenvalues of $\mathbf{J}(\mathbf{x}^a)$ with positive real parts (i.e., the fractional dimension of the unstable manifold). For large $N$, the instability index $\alpha$ depends only on the spectral density $\rho(z; f,\sigma_w)$ and takes the following form:
\begin{equation}
    \label{eq: instability-index}
    \alpha(f) = 
    \begin{cases}
        \frac{f}{\pi}\left(\arccos\sqrt{\frac{f_{\text{st}}}{f}} - \sqrt{\frac{f_{\text{st}}}{f}\left(1-\frac{f_{\text{st}}}{f}\right)}\right) & \text{if } f > f_{\text{st}} \\
        0 & \text{if } f\le f_{\text{st}}
    \end{cases}
\end{equation}
Here, $f_{\text{st}}=\sigma_w^{-2}$ denotes the stability threshold: equilibria with $f\le f_{\text{st}}$ are stable ($\alpha=0$), while those with $f>f_{\text{st}}$ are unstable ($\alpha>0$). 

\subsection{Average over quenched disorders}
The average over Gaussian variables $z_{ij}$ can be performed straightforwardly
\begin{equation}
    \left\langle\exp\Big[\sum_{ij}z_{ij}\Big(\frac{\sigma_w}{\sqrt{N}}\sum_{a=1}^n i\tilde{x}_i^a\phi_j^a\Big)\Big]\right\rangle_{z_{ij}} = \exp\Bigg[-\frac{\sigma_w^2}{2}\sum_{a,b=1}^n\sum_{i=1}^N \tilde{x}_i^a\tilde{x}_i^b \Big(\underbrace{\frac{1}{N}\sum_{j=1}^N\phi_j^a\phi_j^b}_{q^{ab}}\Big)\Bigg]. 
\end{equation}
The {\em overlaps}, $q^{ab}\equiv \frac{1}{N}\sum_{i=1}^N\phi_i^a\phi_i^b$,  measure the scalar products between two equilibria $\mathbf{\phi}^a$, $\mathbf{\phi}^b$ with the same $f$. To decouple different sites, we introduce a set of {\em order parameters} $\{q_{ab}, \mu_a, f_a\}$ as follows 
\begin{gather}
    1 = \int \mathrm{d}q_{ab}\mathrm{d}\hat{q}_{ab} \exp\left[N\hat{q}_{ab}q_{ab} - \sum_{i=1}^N \hat{q}_{ab}\phi_i^a\phi_i^b\right], \\
    1 = \int \mathrm{d}f_a \mathrm{d}\hat{f}_a\exp\left[Nf_a\hat{f}_a - \hat{f}_a\sum_{i=1}^N\Theta_i^a\right], \\
    1 = \int \mathrm{d}\mu_a \mathrm{d}\hat{\mu}_a\exp\left[N\hat{\mu}_a - \hat{\mu}_a\sum_{i=1}^N\phi_i^a + \mathcal{O}(\sqrt{N})\right]. 
\end{gather}
Using these equations in Eq.~\eqref{eq: replicated-kac-rice}, we obtain
\begin{equation}
    \label{eq: replicated-kac-rice-expectation}
    \mathbb{E}[P^n_{\mathbf{w}}]
    =\int \mathrm{d}\theta \mathrm{d}\hat{\theta}\exp\Big[N\Phi_n(\theta,\hat{\theta})\Big].
\end{equation}
Here, $\theta=\{q_{ab},\mu_a,f_a\}$ denotes the set of order parameters, and $\hat{\theta}=\{\hat{q}_{ab}, \hat{\mu}_a, \hat{f}_a\}$ their conjugate variables. The action $\Phi_n(\theta,\hat{\theta})$ is given by 
\begin{equation}
    \label{eq: replicated_action_Phi}
    \Phi_n(\theta,\hat{\theta}) = \sum_{a\le b}\hat{q}_{ab}q_{ab} + \sum_{a}\hat{\mu}_a + \sum_a f_a\hat{f}_a + \sum_a \zeta(f_a,\sigma_w)  + \log \Omega_{n}(\theta,\hat{\theta}), 
\end{equation}
with
\begin{equation}
    \label{eq: replicated_partition_function}
    \begin{aligned}
        \Omega_n(\theta,\hat{\theta}) 
        &= \int_{-\infty}^{\infty}\left(\prod_{a=1}^n \mathrm{d}x_a\right)\exp\left[-\frac{1}{2}(\mathbf{x}-\mathbf{\mu})^T(\sigma_w^2\mathbf{Q})^{-1}(\mathbf{x}-\mathbf{\mu})-\frac{1}{2}\ln\det(2\pi\sigma_w^2\mathbf{Q})\right] \\
        &\quad \times \exp\Big[-\sum_{a\le b}\hat{q}_{ab}\phi_a\phi_b - \sum_a \hat{\mu}_a\phi_a - \sum_a\hat{f}_a\Theta_a\Big]. 
    \end{aligned}
\end{equation}
Here, $\mathbf{x}$ and $\mathbf{\mu}$ are $n$-dimensional vectors in the replica space (each component corresponds to a replica index $a=1,\dots,n$). $\mathbf{Q}$ is the $n\times n$ matrix with elements $q^{ab}$. 

\subsection{Replica symmetric ansatz}
To compute the integral in Eq.~\eqref{eq: replicated_partition_function}, we need to make an explicit assumption about the symmetry of the order parameters under replica permutations. We proceed with the simplest assumption, the {\em replica symmetric (RS) ansatz}, which posits that: 
\begin{gather}
    f_a = f, \\
    \mu_a = \mu, \\
    q_{ab} = \delta_{ab}Q + (1-\delta_{ab})q, 
\end{gather}
and similarly for the conjugate variables (the choice of the minus sign in $\hat{q}$ is for notational convenience)
\begin{gather}
    \hat{f}_a = \hat{f}, \\
    \hat{\mu}_a = \hat{\mu}, \\
    \hat{q}_{ab} = \delta_{ab}\hat{Q} - (1-\delta_{ab})\hat{q}. 
\end{gather}
With the RS symmetry, the partition function $\Omega_n(\theta,\hat{\theta})$ can be written as 
\begin{equation}
    \begin{aligned}
        \Omega_n(\theta,\hat{\theta}) 
        &\propto \int_{-\infty}^{\infty}\left(\prod_{a=1}^n \mathrm{d}x_a\right)
        \exp\left[-\frac{n^2B}{2}\Big(\frac{1}{n}\sum_{a=1}^n x_a-\mu\Big)^2 + \frac{n^2\hat{q}}{2}\Big(\frac{1}{n}\sum_{a=1}^n \phi_a\Big)^2\right]\\
        &\quad\times \prod_{a=1}^n \exp\left[-\frac{A-B}{2}(x_a-\mu)^2-\big(\hat{Q}+\hat{q}/2\big)\phi_a^2-\hat{\mu}\phi_a-\hat{f}\Theta_a\right], 
    \end{aligned}
\end{equation}
where $A$, $B$ denote the diagonal and off-diagonal elements of $(\sigma_w^2\mathbf{Q})^{-1}$, respectively 
\begin{gather}
    A = \frac{Q-q+(n-1)q}{\sigma_w^2(Q-q)\big[Q + (n-1)q\big]}, \\
    B = \frac{-q}{\sigma_w^2(Q-q)\big[Q+(n-1)q\big]}. 
\end{gather}
We introduce two auxiliary variables $y$, $p$ as follows
\begin{gather}
    1 = \int_{-\infty}^{\infty} \mathrm{d}y\mathrm{d}\tilde{y}\, \exp\left[\frac{in\tilde{y}}{\sigma_w^2(Q-q)}\Big(y-\frac{1}{n}\sum_{a=1}^n x_a\Big)\right], \\
    1 = \int_{-\infty}^{\infty} \mathrm{d}p\mathrm{d}\tilde{p}\, \exp\left[in\tilde{p}\Big(p-\frac{1}{n}\sum_{a=1}^n \phi_a\Big)\right],  
\end{gather} 
and plug them into $\Omega_n(\theta,\hat{\theta})$
\begin{equation}
    \begin{aligned}
        \Omega_n(\theta,\hat{\theta}) 
        &\propto \int_{-\infty}^{\infty}\mathrm{d}p\mathrm{d}\tilde{p}\, \exp\left[ \frac{1}{2}n^2\hat{q}p^2 + in\tilde{p}p\right] \\
        &\quad\times \int_{-\infty}^{\infty}\mathrm{d}y\mathrm{d}\tilde{y}\, \exp\left[-\frac{n^2B}{2}(y-\mu)^2 + {in \tilde{y}(y-\mu) \over \sigma_w^2(Q-q)} - {n\tilde{y}^2 \over 2\sigma_w^2(Q-q)}\right] \\
        &\quad \times\prod_{a=1}^n \int_{-\infty}^{\infty} \frac{\mathrm{d}x_a}{\sigma_w\sqrt{2\pi(Q-q)}} \exp\left[-\frac{(x_a-\mu+i\tilde{y})^2}{2\sigma_w^2(Q-q)}  - (\hat{Q}+\hat{q}/2)\phi_a^2 -(\hat{\mu}+i\tilde{p})\phi_a- \hat{f}\Theta_a\right]
    \end{aligned}.
\end{equation}
Integrating out $y$ and $p$ leads to 
\begin{equation}
    \begin{aligned}
        \Omega_n(\theta,\hat{\theta}) 
        &=\int_{-\infty}^{\infty}\mathrm{d}\tilde{y}\mathrm{d}\tilde{p}\,
        \exp\left[\frac{\tilde{y}^2}{2\sigma_w^2q}  - \frac{1}{2}\ln(2\pi\sigma_w^2q) + \frac{\tilde{p}^2}{2\hat{q}} -\frac{1}{2}\ln(2\pi\hat{q})\right] \\
        &\quad\times \prod_{a=1}^n \int_{-\infty}^{\infty}\frac{\mathrm{d}x_a}{\sigma_w\sqrt{2\pi(Q-q)}}
        \exp\left[-\frac{(x_a-\mu +i \tilde{y})^2}{2\sigma_w^2(Q-q)} - (\hat{Q}+\hat{q}/2)\phi_a^2 - (\hat{\mu}+i\tilde{p})\phi_a - \hat{f}\Theta_a \right].  
    \end{aligned}
\end{equation}
$\tilde{y}$, $\tilde{p}$ can be expressed in terms of a pair of independent, standard Gaussian variables $\xi$, $t$ as follows
\begin{gather}
    \tilde{y} = i\sigma_w\sqrt{q}\xi, \\
    \tilde{p} = -it\sqrt{\hat{q}}. 
\end{gather}
Therefore, the partition function $\Omega_n(\theta,\hat{\theta})$ can be expressed as
\begin{equation}
    \Omega_n(\theta,\hat{\theta})
    =\int \mathcal{D}\xi \mathcal{D}t [Z(\xi,t)]^n, 
\end{equation}
where 
\begin{equation}
\label{eq: parfun}
    Z(\xi, t) = \int_{-\infty}^{\infty}\frac{\mathrm{d}x}{\sigma_w\sqrt{2\pi(Q-q)}}
    \exp\left[-\frac{(x-\mu -\xi\sigma_w\sqrt{q})^2}{2\sigma_w^2(Q-q)} - (\hat{Q}+\hat{q}/2)\phi^2 - (\hat{\mu}+t\sqrt{\hat{q}})\phi - \hat{f}\Theta \right]. 
\end{equation}
Using again the replica trick, we have 
\begin{equation}
    \lim_{n\to 0} \frac{1}{n} (\Omega_n-1) = \int\mathcal{D}\xi \mathcal{D}t\; \log Z(\xi, t), 
\end{equation}
thus
\begin{equation}
    \lim_{n\to 0}\frac{1}{n}\log \Omega_{n} 
    =\lim_{n\to 0}\frac{1}{n}\log\left(1 + n\cdot \frac{1}{n}(\Omega_n-1)\right)
    =\int\mathcal{D}\xi \mathcal{D}t \log Z(\xi, t). 
\end{equation}

\subsection{Quenched complexity}

Using the RS ansatz in Eq.~\eqref{eq: replicated-kac-rice-expectation}, we obtain the complexity function 
\begin{equation}
    \mathbb{E}[P^n_{\mathbf{w}}]
    =\int \mathrm{d}\theta \mathrm{d}\hat{\theta}\exp\Big[nN\Sigma_q(\theta,\hat{\theta})\Big]
    \sim \exp\left[nN\Sigma_q^*\right], 
\end{equation}
with
\begin{gather}
    \Sigma_q(\theta,\hat{\theta}) = \hat{Q}Q + \frac{1}{2}\hat{q}q + \hat{\mu} + \hat{f}f + \zeta(f,\sigma_w) + \int \mathcal{D}\xi \mathcal{D}t \log Z(\xi,t). 
\end{gather}
The quenched complexity $\Sigma_q^*$ corresponds to the extremum of $\Sigma_q(\theta,\hat{\theta})$, which is found by solving the saddle-point equations with respect to the order parameters. 

\subsubsection{Saddle-point equations for the single-equilibrium solution ($Q=q$)}
We first consider the case $Q=q$, corresponding to the regime where the network has a unique equilibrium. In this setting, the complexity functional simplifies to 
\begin{equation}
    \Sigma_q(\theta,\hat{\theta})
    =\hat{Q}\big(Q-\langle \phi^2\rangle_{\xi}\big) + \hat{\mu}\big(1-\langle \phi\rangle_{\xi}\big) + \hat{f}\big(f-\langle \Theta\rangle_{\xi}\big) + \frac{\hat{q}}{2}\big(q-\langle\phi^2\rangle_{\xi}\big) + \zeta(f,\sigma_w), 
\end{equation}
where $\phi=\phi(\mu+\xi\sigma_w\sqrt{Q})$ and $\Theta=\Theta(\mu+\xi\sigma_w\sqrt{Q})$ are shorthand notations. The saddle-point equations read
\begin{equation}
    Q = \langle \phi^2\rangle_{\xi}
    = \int \mathcal{D}\xi\; \phi^2(\mu+\xi \sigma_w\sqrt{Q}),
\end{equation}
\begin{equation}
    1 = \langle \phi\rangle_{\xi}
    = \int \mathcal{D}\xi\; \phi(\mu+\xi \sigma_w\sqrt{Q}),
\end{equation}
\begin{equation}
    f = \langle \Theta\rangle_{\xi}
    = \int \mathcal{D}\xi\; \Theta(\mu+\xi \sigma_w\sqrt{Q}).
\end{equation}
These equations admit a unique solution for any value of $\sigma_w$, and coincide with those derived using the cavity method~\cite{Berlemont2022}. While the $Q=q$ solution always exists, it maximizes the function $\Sigma_q(\theta,\hat{\theta})$ only for $\sigma_w\le \sqrt{2}$. For $\sigma_w>\sqrt{2}$, the solution with $Q>q$ dominates, indicating the appearance of exponentially many equilibria, as we will show. 

\subsubsection{Saddle-point equations for the multiple-equilibria solution ($Q>q$)}
When there are multiple equilibria, we define $\sigma=\sigma_w\sqrt{Q-q} > 0$, and rescale the auxiliary variables as 
\begin{gather}
    \hat{Q} \to \sigma^{-2}(\hat{Q}-\hat{q}/2), \\
    \hat{q} \to \sigma^{-2}\hat{q}, \\
    \hat{\mu} = \sigma^{-1}\hat{\mu}. 
\end{gather}
The partition function becomes
\begin{equation}
    Z(\xi, t) 
    =\int_{-\infty}^{\infty}\frac{\mathrm{d}x}{\sqrt{2\pi \sigma^2}} \exp\left[-\frac{1}{2}\left(\frac{x}{\sigma}-\frac{\mu}{\sigma} - \xi \frac{\sigma_w\sqrt{q}}{\sigma}\right)^2 -\hat{Q}\phi(x/\sigma)^2 - (\hat{\mu} + t \sqrt{\hat{q}})\phi(x/\sigma) - \hat{f}\Theta(x/\sigma)\right],
\end{equation}
and the complexity functional reads
\begin{equation}
    \Sigma_q(\theta,\hat{\theta})
    = f\hat{f} + \sigma^{-1}\hat{\mu}  + \sigma^{-2}Q\hat{Q} - \frac{\hat{q}}{2\sigma_w^2} + \zeta(f,\sigma_w) + \int\mathcal{D}\xi \mathcal{D}t \log Z(\xi, t). 
\end{equation}
We introduce the following variables 
\begin{equation}
    m = \frac{\mu}{\sigma}, 
\end{equation}
\begin{equation}
    z = \frac{\sigma_w\sqrt{q}}{\sigma}. 
\end{equation}
With these definitions, the partition function and the complexity functional become
\begin{equation}
    Z(\xi,t)
    =\int_{-\infty}^{\infty} \frac{\mathrm{d}x}{\sqrt{2\pi}}\, \exp\left[-\frac{1}{2}(x-m-\xi z)^2 - \hat{Q}\phi^2(x) - \Big(\hat{\mu} + t \sqrt{\hat{q}}\Big)\phi(x) - \hat{f}\Theta(x)\right], 
\end{equation}
and
\begin{equation}
    \Sigma_q(\theta,\hat{\theta})
    = f\hat{f} + \sigma^{-1}\hat{\mu}  + \frac{\hat{Q}}{\sigma_w^2}(1+z^2) - \frac{\hat{q}}{2\sigma_w^2} + \zeta(f,\sigma_w) + \int\mathcal{D}\xi \mathcal{D}t\; \log Z(\xi, t). 
\end{equation}
We define the normalized probability distribution $p(x | \xi, t)$ as 
\begin{equation}
    p(x | \xi, t) 
    \propto  \exp\left[-\frac{1}{2}(x-m-\xi z)^2 - \hat{Q}\phi^2 - \Big(\hat{\mu} + t \sqrt{\hat{q}}\Big)\phi - \hat{f}\Theta(x)\right], 
\end{equation}
and denote integration over $x$, $\xi$ and $t$ with respect to this distribution as $\langle \cdot \rangle$, i.e., 
\begin{equation}
    \langle O(x | \xi,t)\rangle = \int\mathcal{D}\xi \mathcal{D}t \int_{-\infty}^{\infty}\mathrm{d}x\, p(x|\xi, t)O(x,\xi,t). 
\end{equation}
The saddle-point equations are obtained by taking derivatives of $\Sigma_q$ with respect to the rescaled variables $\theta_{new} = \{\hat{f}, \hat{Q}, \hat{\mu}, \hat{q}, m, \sigma, z\}$. For $\sigma$, $\hat{\mu}$, $\hat{f}$, we have
\begin{gather}
    \label{eq: quenched_saddle_sigma}
    \hat{\mu} = 0. \\
    \label{eq: quenched_saddle_muhat}
    \sigma^{-1} = \langle \phi\rangle. \\
    \label{eq: quenched_saddle_fhat}
    f = \langle \Theta\rangle.
\end{gather}
For $\hat{Q}$, $\hat{q}$, $m$, $z$, the saddle-point equations read (where we used the fact that $\langle \xi\rangle=0$ and $\langle \xi^2\rangle=1$)
\begin{gather}
    \label{eq: quenched_saddle_Qhat}
    \frac{1+z^2}{\sigma_w^2} = \langle \phi^2\rangle. \\
    \label{eq: quenched_saddle_v}
    \frac{\hat{q}}{\sigma_w^2} + \langle t \phi\rangle = 0. \\
    \label{eq: quenched_saddle_m}
    m - \langle x\rangle = 0. \\
    \label{eq: quenched_saddle_z}
    \frac{2\hat{Q}z}{\sigma_w^2} + \langle\xi x\rangle - z = 0. 
\end{gather}
Eqs.~(\ref{eq: quenched_saddle_Qhat}-\ref{eq: quenched_saddle_z}) can be solved self-consistently for a given $\hat{f}$, then we substitute the solutions into Eqs.~(\ref{eq: quenched_saddle_muhat}, \ref{eq: quenched_saddle_fhat}) to determine the remaining parameters. 

\subsubsection{Finite-size correction for the quenched complexity}
We consider the leading-order correction to the quenched complexity by imposing an $N$-dependent balance condition
\begin{equation}
    \langle \phi\rangle = 1 \quad \Rightarrow \quad \langle \phi\rangle = 1-\frac{\mu}{\sqrt{N}}. 
\end{equation}
This modification is equivalent to replacing the term $\hat{\mu}\cdot 1$ in the complexity function with $\hat{\mu}(1-\frac{\mu}{\sqrt{N}})$. Expressed in terms of the rescaled variables, the complexity becomes
\begin{equation}
    \Sigma_q(\theta,\hat{\theta})
    = \Sigma_q^{(0)}(\theta, \hat{\theta}) - \frac{m}{\sqrt{N}}\hat{\mu}  + \mathcal{O}(N^{-1}),
\end{equation}
where $\Sigma_q^{(0)}(\theta, \hat{\theta})$ denotes the asymptotic complexity. This correction affects only the saddle-point equation with respect to $\hat{\mu}$, which is modified to
\begin{equation}
    \sigma^{-1} = \langle \phi\rangle + \frac{m}{\sqrt{N}}, 
\end{equation}
while all other equations remain unchanged (the equation for $m$ is unaffected, since the saddle-point conditions still enforce $\hat{\mu}=0$). 

Because both $\langle\phi\rangle$ and $m$ can be determined independently of $\sigma$, their values coincide with those obtained in the limit $N\to\infty$. Using the definitions of $m$ and $\sigma$, we obtain the following finite-size corrections
\begin{gather}
    \label{eq: finite-size-Q}
    Q_{N} = Q_{\infty}\left(1 + \frac{\mu_{\infty}}{\sqrt{N}}\right)^{-2}, \\
    \label{eq: finite-size-q}
    q_{N} = q_{\infty}\left(1 + \frac{\mu_{\infty}}{\sqrt{N}}\right)^{-2}, \\ 
    \label{eq: finite-size-mu}
    \mu_{N} = \mu_{\infty}\left(1 + \frac{\mu_{\infty}}{\sqrt{N}}\right)^{-1}. 
\end{gather}

\subsection{Annealed complexity}
The average number of equilibria, $\mathbb{E}[P_{\mathbf{w}}]$, can be estimated by setting $n=1$ in Eq.~\eqref{eq: replicated_partition_function}. Using the saddle-point approximation, we have 
\begin{equation}
    \lim_{N\to\infty}\frac{1}{N}\log \mathbb{E}[P^{n=1}_{\mathbf{w}}]
    =\underset{\theta,\hat{\theta}}{\text{argmax}}\; \Sigma_a(\theta, \hat{\theta}), 
\end{equation}
where $\Sigma_a(\theta,\hat{\theta})$ is the corresponding functional for the annealed complexity
\begin{equation}
    \label{eq: annealed_complexity}
    \Sigma_{a}(\theta,\hat{\theta}) = \hat{Q}Q  + \hat{\mu} + \hat{f}f + \zeta(f,\sigma_w) + \ln \int_{-\infty}^{\infty}\frac{\mathrm{d}x}{\sqrt{2\pi\sigma_w^2 Q}}\exp\left[-\frac{(x-\mu)^2}{2\sigma_w^2Q}-\hat{Q}\phi^2 - \hat{\mu}\phi - \hat{f}\Theta\right]. 
\end{equation}
The parameters $\theta$ and $\hat{\theta}$ can be determined by solving the corresponding saddle-point equations. As a first step, we evaluate the Gaussian integral for $x<0$ in the last term
\begin{equation}
    Z_- 
    = \int_{-\infty}^{0} \frac{\mathrm{d}x}{\sqrt{2\pi\sigma_w^2Q}} \exp\left[-\frac{(x-\mu)^2}{2\sigma_w^2Q}\right]
    = \frac{1}{2}\text{erfc}(z),\quad z\equiv \frac{\mu}{\sqrt{2\sigma_w^2Q}}. 
\end{equation}
Similarly, for $x>0$ we obtain
\begin{equation}
    \begin{aligned}
        Z_+
        &= \int_{0}^{\infty} \frac{\mathrm{d}x}{\sqrt{2\pi\sigma_w^2Q}} \exp\left[-\frac{(x-\mu)^2}{2\sigma_w^2Q}-\hat{\mu}x - \hat{Q}x^2 - \hat{f}\right] \\
        & = \int_{0}^{\infty} \frac{\mathrm{d}x}{\sqrt{2\pi\sigma_w^2Q}} \exp\left[-\beta(x-\alpha)^2 + \beta\alpha^2 - z^2 - \hat{f}\right] \\
        &=\frac{\text{erfc}(y)}{2\sqrt{2\sigma_w^2Q\beta}}\exp\left[y^2-z^2-\hat{f}\right], \quad y\equiv -\alpha\sqrt{\beta}, 
    \end{aligned}
\end{equation}
where $\beta$ and $\alpha$ are defined by completing the square in the exponent for $x>0$
\begin{equation}
    \beta = \frac{1}{2\sigma_w^2Q} + \hat{Q}, \quad 
    \alpha = \frac{1}{2\beta}\left(\frac{\mu}{\sigma_w^2Q}-\hat{\mu}\right). 
\end{equation}
The saddle-point equation with respect to $\hat{f}$ gives
\begin{equation}
    f - \frac{Z_+}{Z_- + Z_+} = 0 
    \quad \Rightarrow\quad  \frac{Z_+}{Z_-} = \frac{f}{1-f} = \frac{\text{erfc}(y)}{\text{erfc}(z)}\frac{\exp\big[y^2-z^2-\hat{f}\big]}{\sqrt{2\sigma_w^2Q\beta}}. 
\end{equation}
Hence, $\hat{f}$ is uniquely determined by the remaining parameters. Substituting the solution for $\hat{f}$ allows us to split the logarithm term 
\begin{equation}
    \begin{aligned}
        \ln(Z_- + Z_+) 
        &=\ln\left[(1-f)\cdot\frac{Z_-}{1-f} + f\cdot\frac{Z_+}{f}\right] \\
        &=(1-f)\ln Z_- + f \ln Z_+ - f\ln f - (1-f)\ln(1-f). 
    \end{aligned}
\end{equation}
Substituting $\hat{Q} \to \beta - \frac{1}{2\sigma_w^2Q}$ and $\hat{\mu} \to \frac{\sqrt{2}z}{\sqrt{\sigma_w^2Q}}+2\sqrt{\beta}y$ yields the complexity function
\begin{equation}
    \begin{aligned}
        \Sigma_a(y,z,\beta, Q)
        &= (1-f)\ln\text{erfc}(z) - fz^2 + \frac{2z}{\sqrt{2\sigma_w^2Q}} + f \ln\text{erfc}(y) + fy^2 + 2\sqrt{\beta} y \\
        &\quad +  \beta Q - \frac{f}{2}\ln(2\sigma_w^2Q\beta) + \zeta(f,\sigma_w) -\frac{1}{2\sigma_w^2} - f\ln f - (1-f)\ln (1-f) - \ln 2,
    \end{aligned}
\end{equation}
We introduce the function 
\begin{equation}
    \psi(x) = \frac{\exp(-x^2)}{\sqrt{\pi}\text{erfc}(x)} = -\frac{1}{2}\frac{\mathrm{d}}{\mathrm{d}x}\ln \text{erfc}(x), 
\end{equation}
and the saddle-point equations with respect to $z$, $y$, $Q$ and $\beta$ are given by 
\begin{gather}
    \label{eq: annealed_saddle_1}
    f z + (1-f)\psi(z) = \frac{1}{\sqrt{2\sigma_w^2Q}}, \\
    \label{eq: annealed_saddle_2}
    f(\psi(y)-y) = \sqrt{\beta}, \\
    \label{eq: annealed_saddle_3}
    - \frac{2z}{\sqrt{2\sigma_w^2Q}} + 2\beta Q - f = 0, \\
    \label{eq: annealed_saddle_4}
    2\sqrt{\beta}y+2\beta Q - f = 0, 
\end{gather}
which can be summarized into one equation 
\begin{equation}
    \label{eq: annealed_saddle_5}
    f-2\beta Q = 2fy(\psi(y)-y) = -2z(fz + (1-f)\psi(z)). 
\end{equation}
We observe that $2y(\psi(y)-y)$ is a monotonically increasing function with range $(-\infty, 1)$. Thus, for a given $f$ and $\lambda\equiv 2\beta Q>0$, $y$ is uniquely determined as a function of $\lambda$; we denote this solution by $y_{\lambda}$. From Eqs.~(\ref{eq: annealed_saddle_2}, \ref{eq: annealed_saddle_3}), $z_{\lambda}$ can also be computed as a function of $\lambda$
\begin{equation}
    \label{eq: annealed_saddle_6}
    z_{\lambda} = -\sigma_w y_{\lambda}\sqrt{\lambda}. 
\end{equation}
Substituting Eq.~\eqref{eq: annealed_saddle_6} into Eq.~\eqref{eq: annealed_saddle_5} yields the final equation for $\lambda$
\begin{equation}
    f - \lambda + 2z_{\lambda}\big[fz_{\lambda}+(1-f)\psi(z_{\lambda})\big] = 0. 
\end{equation}

\subsubsection{Finite-size correction for the annealed complexity}
The finite-size correction to the annealed complexity is obtained in the same way as for the quenched case. We replace $\hat{\mu}$ with $\hat{\mu}(1-\frac{\mu}{\sqrt{N}})$ in Eq.~\eqref{eq: annealed_complexity} and obtain the following modified saddle-point equations
\begin{gather}
    \label{eq: annealed_modified_saddle_z}
    f z + (1-f)\psi(z) + \frac{2z}{\sqrt{N}} + \frac{y}{\sqrt{N}}\sqrt{2\sigma_w^2Q\beta} = \frac{1}{\sqrt{2\sigma_w^2Q}}, \\
    \label{eq: annealed_modified_saddle_y}
    f(\psi(y)-y)+\frac{z}{\sqrt{N}}\sqrt{2\sigma_w^2Q\beta} = \sqrt{\beta}, \\
    \label{eq: annealed_modified_saddle_Q}
    2\beta Q - f - \frac{2z}{\sqrt{2\sigma_w^2Q}} - \frac{2yz}{\sqrt{N}}\sqrt{2\sigma_w^2Q\beta} = 0, \\
    \label{eq: annealed_modified_saddle_beta}
    2\beta Q - f + 2\sqrt{\beta}y - \frac{2yz}{\sqrt{N}}\sqrt{2\sigma_w^2Q\beta} = 0. 
\end{gather} 
From the last two equations we obtain
\begin{equation}
    \label{eq: annealed_saddle_zy}
    \frac{z}{y} = -\sigma_w\sqrt{\lambda}, 
\end{equation}
which implies
\begin{equation}
    \hat{\mu} = \frac{\sqrt{2}z}{\sqrt{\sigma_w^2Q}}+2\sqrt{\beta}y = 0. 
\end{equation}
The modified saddle-point equations yield the same solutions for $y_{\lambda}$ and $z_{\lambda}$ as functions of $\lambda$. Moreover, the annealed complexity itself remains unaffected by the finite-size correction (since $\hat{\mu}=0$). As in the quenched case, finite-size effects appear only in the geometric parameters $Q$ and $\mu$, which scale according to Eqs.~(\ref{eq: finite-size-Q}, \ref{eq: finite-size-mu}). 

\section{Geometry of the equilibria}

Here we discuss the organization of the equilibria in the phase space, i.e., their geometry, as revealed by the replica-symmetric theory presented in the previous section. The theory determines a set of `order parameters' as a function of the fraction of active neurons $f$ (and the synaptic gain $\sigma_w$). Among these,  $Q\left(f\right)$ and $q\left(f\right)$ have a simple geometrical interpretation. The distance of $\boldsymbol{\phi}^a$ (with $f^a=f$) from the origin (i.e., the norm of $\boldsymbol{\phi}^a$) is given by 

\begin{equation}\label{eq:def_Da}
    D^a\left(f\right)=\sqrt{\sum_{j=1}^N\left(\phi^a_j-0\right)^2}=\sqrt{\frac{N}{N}\sum_{j=1}^N\left(\phi^a_j\right)^2}=\sqrt{NQ\left(f\right)}\equiv D\left(f\right)
\end{equation}

\noindent which does not depend on $a$ in the limit $N\to\infty$. Therefore, all $\boldsymbol{\phi}^a$'s lay on the hypersphere centered at the origin with radius $D\left(f\right)$, in the non-negative orthant, because $\phi^a_j\geq0$ for all $a$ and $j$ by definition. The radius of the hypersphere decreases with $f$; the larger $f$, the closer to the origin the corresponding $\boldsymbol{\phi}^a$'s.

The angle, $\theta^{ab}\left(f\right)$, between two distinct $\boldsymbol{\phi}^a$ and $\boldsymbol{\phi}^b$ (i.e., $a\neq b$) with $f^a=f^b=f$, is given by

\begin{equation}\label{eq:actual_angle}
    \cos\theta^{ab}\left(f\right)=\frac{\sum_{j=1}^N\phi^a_j\phi^b_j}{\sqrt{\sum_{j=1}^N\left(\phi^a_j\right)^2\sum_{k=1}^N\left(\phi^b_k\right)^2}}=\frac{Nq\left(f\right)}{\sqrt{NQ\left(f\right)NQ\left(f\right)}}=\frac{q\left(f\right)}{Q\left(f\right)}\equiv\cos\theta\left(f\right)
\end{equation}

\noindent which, again, does not depend on the specific solutions $a$ and $b$ chosen, but only on $f$. Therefore, every pair of distinct solutions, $\boldsymbol{\phi}^a$ and $\boldsymbol{\phi}^b$, forms the same angle (at parity of $f$). This angle also decreases with $f$; the larger $f$, the smaller the angle $\theta(f)$, and the larger the cosine similarity, i.e., $\cos\theta\left(f\right)$, between the corresponding $\boldsymbol{\phi}^a$'s.

The fact that every pair of distinct solutions makes the same angle is not surprising in high-dimensional spaces; it is just a manifestation of the law of large numbers. To see this, let us consider the constraints that any solution $\boldsymbol{\phi}^a$ with $f^a=f$ has to satisfy (in the limit $N\to\infty$). There are three constraints; they are

\begin{equation}\label{eq:constr_1}
    \phi^a_j\geq0\;\;\mbox{for all $j$},
\end{equation}

\noindent because $\phi\left(x\right)=x\Theta\left(x\right)$,

\begin{equation}\label{eq:constr_2}
    \frac{1}{N}\sum_i\phi^a_i=1,
\end{equation}

\noindent which is a consequence of the balanced regime our model network operates in, and, finally,

\begin{equation}\label{eq:constr_3}
    \frac{1}{N}\sum_i\left(\phi^a_i\right)^2=Q\left(f\right),
\end{equation}

\noindent which fixes the norm of the $\boldsymbol{\phi}^a$. What is the angle formed by two vectors, $\boldsymbol{\hat{\phi}}^a$ and $\boldsymbol{\hat{\phi}}^b$, whose elements are randomly (according to some distribution) and {\em independently} chosen such that the above constraints are satisfied? To answer the question, we evaluate their scalar product. This is given by

\begin{equation}
\boldsymbol{\hat{\phi}}^a\cdot\boldsymbol{\hat{\phi}}^b=\sum_i\hat{\phi}^a_i\hat{\phi}^b_i=Nf^2\langle\hat{\phi}^a_i\rangle_A\langle\hat{\phi}^b_i\rangle_A=Nf^2\frac{1}{f}\frac{1}{f}=N
\end{equation}

\noindent because the probability that both $\hat{\phi}^a_i$ and $\hat{\phi}^b_i$ are different from zero is $f^2$, and the average of $\hat{\phi}^a_i$ restricted to the `active' neurons, that we have denoted $\langle\hat{\phi}^a_i\rangle_A$ in the above equation, is $1/f$ (because of Eq.~\eqref{eq:constr_2}). Thus, the angle they form, $\hat{\theta}\left(f\right)$, is given by

\begin{equation}\label{eq:random_angle}
    \cos\hat{\theta}\left(f\right)=\frac{\boldsymbol{\hat{\phi}}^a\cdot\boldsymbol{\hat{\phi}}^b}{\lVert\boldsymbol{\hat{\phi}}^a\rVert\lVert\boldsymbol{\hat{\phi}}^b\rVert}=\frac{N}{\sqrt{NQ\left(f\right)}\sqrt{NQ\left(f\right)}}=\frac{1}{Q\left(f\right)}
\end{equation}

\noindent and it is independent of $\boldsymbol{\hat{\phi}}^a$ and $\boldsymbol{\hat{\phi}}^b$. Comparing Eq.~\eqref{eq:random_angle} with Eq.~\eqref{eq:actual_angle}, and noting that $q\left(f\right)>1$, we see that $\theta\left(f\right)<\hat{\theta}(f)$. In other words, actual equilibria are more similar, as measured by cosine similarity, than random vectors satisfying the same constraints, that is, actual equilibria are significantly correlated. 

The origin of these correlations should be obvious; they result from the fact that $\boldsymbol{\phi}^a$'s are all equilibria of the {\em same} sample network. In fact, $x^a_i$ and $x^b_i$, and hence $\phi^a_i$ and $\phi^b_i$ (i.e., same neuron, different equilibria), are correlated because the $w_{ij}$'s do not change with the equilibria (i.e., the $w_{ij}$'s are quenched variables). These correlations are quantified by $q\left(f\right)$ as discussed above, and they are only accessible through the quenched theory; in the annealed theory there is no $q\left(f\right)$, because of the {\em direct} averaging over the $w_{ij}$, which effectively removes the correlations. Correlations between $\phi^a_i$ and $\phi^a_j$ (i.e., same equilibrium, different neurons) {\em individually} got to $0$ in the limit $N\to\infty$, because of the mean-field nature of our model, both in the quenched and the annealed theory. As we will see below, these correlations also contain geometrical information. Unfortunately, however, they are more difficult to describe analytically. 

To better understand the spatial organization of the equilibria, let us consider a (very) large sample network. For any $f$ such that $\Sigma_q\left(f\right)>0$, we can define the {\em center-of-mass} of the corresponding $\boldsymbol{\phi}^a$'s (i.e., with $f^a=f$) as

\begin{equation}\label{eq:sm_restricted_average}
    \overline{\boldsymbol{\phi}}\left(f\right)=\frac{1}{P_\mathbf{w}\left(f\right)}\sum_{\substack{a\vert f^a=f}}\boldsymbol{\phi}^a
\end{equation}

\noindent where $P_\mathbf{w}\left(f\right)$ is the number of equilibria with $f^a=f$ and the sum is restricted to the equilibria with $f^a=f$. We note that $\overline{\boldsymbol{\phi}}\left(f\right)$ depends on the specific realization of the connectivity matrix $\mathbf{w}$.

To determine the angle, $\omega^a\left(f\right)$, between $\boldsymbol{\phi}^a$ and $\overline{\boldsymbol{\phi}}\left(f\right)$, we evaluate their scalar product. This is given by

\begin{align}
\boldsymbol{\phi}^a\cdot\overline{\boldsymbol{\phi}}\left(f\right) & =\lVert\boldsymbol{\phi}^a\rVert\lVert\overline{\boldsymbol{\phi}}\left(f\right)\rVert\cos\omega^a\left(f\right)=\sum_i\phi^a_i\frac{1}{P_\mathbf{w}\left(f\right)}\sum_{\substack{b\vert f^b=f}}\phi^b_i=\frac{1}{P_\mathbf{w}\left(f\right)}\sum_{\substack{b\vert f^b=f}}\sum_i\phi^a_i\phi^b_i \\ & = \frac{N}{P_\mathbf{w}\left(f\right)}\left(Q\left(f\right)+\left(P_\mathbf{w}\left(f\right)-1\right)q\left(f\right)\right)=Nq\left(f\right);\;\;\left(N\to\infty\right)
\end{align}

\noindent The norm of $\boldsymbol{\phi}^a$ is simply $\lVert\boldsymbol{\phi}^a\rVert=\sqrt{NQ\left(f\right)}$; the norm of $\overline{\boldsymbol{\phi}}\left(f\right)$ is

\begin{align}
    \lVert\overline{\boldsymbol{\phi}}\left(f\right)\rVert^2 & =\sum_i\left(\frac{1}{P_\mathbf{w}\left(f\right)}\sum_{\substack{a\vert f^a=f}}\phi^a_i\right)^2=\frac{1}{P_\mathbf{w}^2\left(f\right)}\sum_{\substack{a,b\\ f^a=f^b=f}}\sum_i\phi^a_i\phi^b_i \\ & =\frac{N}{P_\mathbf{w}^2\left(f\right)}\left(P_\mathbf{w}\left(f\right)Q\left(f\right)+P_\mathbf{w}\left(f\right)\left(P_\mathbf{w}\left(f\right)-1\right)q\left(f\right)\right)=Nq\left(f\right);\;\;\left(N\to\infty\right)
\end{align}

\noindent and, putting all together, we find

\begin{equation}
    \cos\omega^a\left(f\right)=\sqrt{\frac{q\left(f\right)}{Q\left(f\right)}}\equiv\cos\omega\left(f\right)
\end{equation}

\noindent which does not depend on $a$, nor on $\mathbf{w}$, because both $Q\left(f\right)$ and $q\left(f\right)$ are self-averaging. Therefore, all $\boldsymbol{\phi}^a$'s with $f^a=f$ form the same angle with $\overline{\boldsymbol{\phi}}\left(f\right)$. This implies that the $\boldsymbol{\phi}^a$'s are contained in a spherical cap, whose pole is determined by the intersection, in the non-negative orthant, between the straight line through the origin defined by $\overline{\boldsymbol{\phi}}\left(f\right)$, and the hypersphere of radius $D\left(f\right)=\sqrt{NQ\left(f\right)}$ centered at the origin; the polar angle is given by $\omega\left(f\right)$. In fact, in the limit $N\to\infty$, the volume of the cap is concentrated in the rim, and any vector between the origin and a point in the rim forms the same angle with $\overline{\boldsymbol{\phi}}\left(f\right)$, that is, $\omega\left(f\right)$. Note that this does not imply that all such vectors are equilibria, but rather that all the equilibria $\boldsymbol{\phi}^a$ (with $f^a=f$) are {\em inside} the rim.    

For `random' equilibria (as defined above), the polar angle, $\hat{\omega}\left(f\right)$, is

\begin{equation}
    \cos\hat{\omega}\left(f\right)=\sqrt{\frac{1}{Q\left(f\right)}}<\cos\omega\left(f\right)
\end{equation}

\noindent and, hence, $\hat{\omega}\left(f\right)>\omega\left(f\right)$ for all $f$. Thus, actual equilibria can be enclosed in a volume (much) smaller than the volume needed to enclose `random' equilibria. This is another manifestation of the fact that actual equilibria are correlated.

The above arguments suggest that we also should expect significant correlations between equilibria with different fractions of active neurons. These can be quantified by introducing the overlaps, $q^{ab}\left(f_1,f_2\right)$, between two equilibria, $\boldsymbol{\phi}^a$ with $f^a=f_1$ and $\boldsymbol{\phi}^b$ with $f^b=f_2$. These are defined by

\begin{equation}\label{eq:sm_overlap_f1f2}
    q^{ab}\left(f_1,f_2\right)=q\left(f_1,f_2\right)\equiv\frac{1}{N}\sum_{i}\phi^a_i\phi^b_i
\end{equation}

\noindent where we have dropped the dependence on the pair of solutions, because we expect $q^{ab}\left(f_1,f_2\right)$ to be self-averaging. In fact, this can be seen more `rigorously', and $q\left(f_1,f_2\right)$ can be computed analytically, by a relatively straightforward extension of the (replica-symmetric) theory presented in the previous section. 

For `random' equilibria, $q\left(f_1,f_2\right)=1$. In fact, the probability that both $\phi^a_i$ and $\phi^b_i$ are different from zero is $f_1f_2$, while $\langle\phi^a_i\rangle_A=1/f_1$ and $\langle\phi^b_i\rangle_A=1/f_2$. Thus, $q\left(f_1,f_2\right)>1$ indicates significant pair-wise correlations among equilibria with two different fractions, $f_1$ and $f_2$, of active neurons. 

As discussed above, both the $\boldsymbol{\phi}^a$'s and the $\boldsymbol{\phi}^b$'s are inside spherical caps. The relative position of these spherical caps can be conveniently quantified by the cosine similarity between the corresponding polar vectors; this is given by

\begin{equation}
    \cos\omega\left(f_1,f_2\right)\equiv\frac{q\left(f_1,f_2\right)}{\sqrt{q\left(f_1\right)q\left(f_2\right)}}
\end{equation}

\noindent where $\omega\left(f_1,f_2\right)$ is the angle between the center-of-mass of the $\boldsymbol{\phi}^a$'s with $f^a=f_1$, $\overline{\boldsymbol{\phi}}\left(f_1\right)$, and the center-of-mass of the $\boldsymbol{\phi}^b$'s with $f^b=f_2$, $\overline{\boldsymbol{\phi}}\left(f_2\right)$.

\section{Dynamical mean-field theory} \label{sec:dmft}

Here we summarize the dynamical mean-field theory (DMFT) for our model network. Our presentation is short and {\em ad rem}. Detailed derivations of the DMFT can be found in the literature; a very detailed and pedagogical one is \cite{crisanti2018path}.

We are concerned only with the steady state of the dynamics. We define

\begin{equation}
    \Delta\left(\tau\right)\equiv\frac{1}{N}\sum_{i=1}^N\left(x_i\left(t\right)-\mu\right)\left(x_i\left(t+\tau\right)-\mu\right)
\end{equation}

\noindent where $\mu\equiv N^{-1}\sum_{i=1}^Nx_i\left(t\right)$, which is constant in time (at equilibrium), and

\begin{equation}
    q\left(\tau\right)\equiv\frac{1}{N}\sum_{i=1}^N\phi_i\left(t\right)\phi_i\left(t+\tau\right)
\end{equation}

\noindent Clearly, $\Delta\left(\tau\right)=\Delta\left(-\tau\right)$ and $q\left(\tau\right)=q\left(-\tau\right)$. For later convenience, we introduce the following short-hand notations: $\Delta_0\equiv\Delta\left(0\right)$, $\Delta_\infty\equiv\lim_{\tau\to\infty}\Delta\left(\tau\right)$ and, analogously, for $q_0$ and $q_\infty$ (corresponding to $Q_\text{D}$ and $q_\text{D}$ in the main text). Obviously, $\Delta_0\geq\Delta_\infty$ and $q_0\geq q_\infty$.

Two self-consistency conditions can be immediately derived. From the balance condition, we obtain
\begin{equation}\label{eq:dmft_eq1}
    1=\int\mathrm{D}z\;\phi\left(\mu+z\sqrt{\Delta_0}\right)
\end{equation}

\noindent which expresses the fact that the average firing rate in the network is constant in time and determined by the balance condition. In the limit $\tau\to\infty$, $x_i\left(t\right)$ and $x_i\left(t+\tau\right)$ become uncorrelated. Thus,

\begin{equation}\label{eq:dmft_eq2}
    \Delta_\infty=\sigma_w^2\int\mathrm{D}z\;\left(\int\mathrm{D}y\;\phi\left(\mu+y\sqrt{\Delta_0-\Delta_\infty}+z\sqrt{\Delta_\infty}\right)\right)^2,
\end{equation}

\noindent where $y$ and $z$ are independent standard Gaussian variables. We note that when the network dynamics is in an equilibrium, i.e., $\Delta_0=\Delta_\infty$ and, hence, $q_0=q_\infty$. In the general case, however, we need a third self-consistent equation that has no (obvious) counterpart in the `static' theory.

The third self-consistent condition can be obtained as follows: $\Delta\left(\tau\right)$ and $q\left(\tau\right)$ are related by the following equation

\begin{equation}\label{eq:dmft_eq3}
    \ddot{\Delta}=\Delta-\sigma_w^2q=-\frac{\partial V}{\partial\Delta}
\end{equation}

\noindent where the double dot denotes the second derivative with respect to time (i.e., $\tau$) and $V\left(\Delta\right)$ is a `potential', which can always be defined because the dynamics is $1$-D. Clearly, the solutions of Eq.~\eqref{eq:dmft_eq3} conserve the total `energy', i.e.,

\begin{equation}\label{eq:dmft_eq4}
    \frac{1}{2}\dot{\Delta}^2+V\left(\Delta\right)=\mathrm{constant}
\end{equation}

\noindent along all the orbits of Eq.~\eqref{eq:dmft_eq3}. In the limit $\tau\to\infty$, $\dot{\Delta}(\tau)=0$. Also, $\Delta(\tau)$ is symmetric around $0$ and reaches its maximum value there, hence $\dot{\Delta}(0)=0$. Using this information into Eq.~\eqref{eq:dmft_eq4}, we obtain

\begin{equation}\label{eq:dmft_eq5}
    V\left(\Delta_0\right)=V\left(\Delta_\infty\right)
\end{equation}

\noindent because the `kinetic energy' is $0$ both at $\tau=0$ and at $\tau\to\infty$. Eq.~\eqref{eq:dmft_eq5} provides the required third self-consistent condition. The potential $V$ can be explicitly computed (see \cite{crisanti2018path}); in our case, it reads

\begin{equation}\label{eq:dmft_eq6}
    V\left(\Delta\right)=-\frac{\Delta^2}{2}+\sigma_w^2\int\mathrm{D}z\;\left(\int\mathrm{D}y\;\Phi\left(\mu+y\sqrt{\Delta_0-\Delta}+z\sqrt{\Delta}\right)\right)^2
\end{equation}

\noindent where $\Phi\left(x\right)$ is the anti-derivative of $\phi\left(x\right)$. In our case,

\begin{equation}
    \Phi\left(x\right)=\frac{x^2}{2}\Theta\left(x\right)
\end{equation}

\section{Bounding the Lyapunov spectrum}

\begin{figure}[tb]
    \centering
    \includegraphics[width=0.95\textwidth]{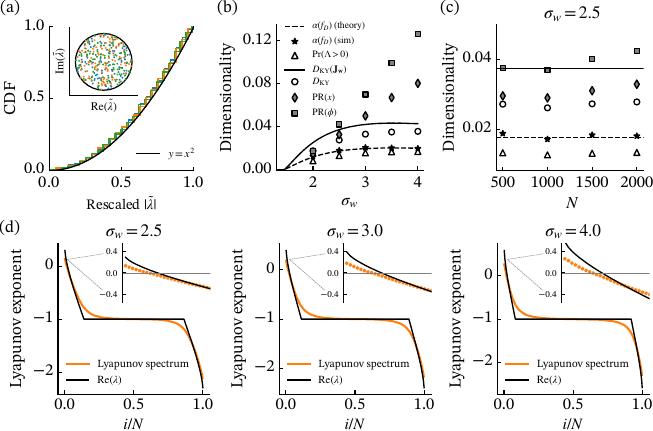}
    \caption{
    (a) Spectral distribution of $\tilde{\lambda}\equiv(\sigma_w\sqrt{f})^{-1}(1+\lambda)$, where $\lambda$ is a non-trivial eigenvalue of the Jacobian computed at points sampled from the attractor. $N=2000$, $\sigma_w=2.5$. 
    (b) Comparison of different estimates of the fractional dimension of the attractor. Symbols denote the instability index $\alpha(f_D)$, fraction of positive Lyapunov exponents $\Pr(\Lambda>0)$, the Kaplan-Yorke dimension ($D_{\text{KY}}$), and the participation ratios computed using $\mathbf{x}$ or $\phi(\mathbf{x})$ ($\text{PR}(x)$ and $\text{PR}(\phi)$, respectively). 
    (c) Fractional dimension estimates as a function of $N$ for $\sigma_w=2.5$. 
    (d) Real parts of the eigenvalues of $\mathbf{J}_{\mathbf{w}}(\tilde{\mathbf{x}})$ compared to the full Lyapunov spectra for several values of $\sigma_w$; $N=500$. Inset: largest 5\% of exponents/eigenvalues. 
    }
    \label{supfig4}
\end{figure}

In this Section, we provide a theoretical rationale for the observation that the real parts of the Jacobian spectrum on the attractor provide a fairly accurate approximation to the Lyapunov exponents (LEs); see Fig.~\ref{supfig4}. The following discussion is not intended to provide a rigorous mathematical proof; rather, it offers a heuristic justification in the tradition of theoretical physics. Nevertheless, we believe that many, if not all, of our arguments can be made mathematically precise.

Let us start by briefly introducing LEs; we refer the interested reader to \cite{eckmann1985ergodic,dorfman1999introduction} for a more thorough presentation. LEs are a standard tool to quantify sensitive dependence on initial conditions and, by extension, the presence of chaos in dynamical systems. The basic idea is very simple. One considers an {\em unperturbed} trajectory passing through $\mathbf{x}_0\equiv\mathbf{x}\left(0\right)$ at time $t=0$, alongside a {\em perturbed} trajectory passing through $\mathbf{x}_0+\delta\mathbf{x}_0$ at the same time, and asks how different the two trajectories will be in the future. After a time $t>0$,
\begin{equation}
\mathbf{x}\left(t\right)=\Phi^t\left(\mathbf{x}_0\right)\;\;\;\mbox{and}\;\;\;\mathbf{x}\left(t\right)+\delta\mathbf{x}\left(t\right)=\Phi^t\left(\mathbf{x}_0+\delta\mathbf{x}_0\right)
\end{equation}

\noindent where the flow $\Phi^t\left(\cdot\right)$, mapping the initial condition onto the system's state at time $t$, is obtained by integrating the system's dynamical equations (i.e., in our case, $\dot{\mathbf{x}}=\mathbf{v}_\mathbf{w}\left(\mathbf{x}\right)$) and $\delta\mathbf{x}\left(t\right)$ quantifies the difference between the two trajectories after a time $t$. Then, if for all sufficiently small $\delta\mathbf{x}_0$, $\lVert\delta\mathbf{x}\left(t\right)\rVert\to0$ for $t\to+\infty$, the orbit through $\mathbf{x}_0$ is asymptotically stable. In other words, suitably small differences in the initial conditions lead to the same dynamics (i.e., asymptotic convergence to the `same' orbit). On the other hand, if $\lVert\delta\mathbf{x}\left(t\right)\rVert$ for $t\to+\infty$ cannot be made arbitrarily small by reducing $\lVert\delta\mathbf{x}_0\rVert$, the orbit through $\mathbf{x}_0$ is unstable. In other words, arbitrarily small differences in the initial conditions lead to different dynamics (i.e., to macroscopically distinct orbits). Clearly, there is a sensitive dependence on the initial conditions in this case.

For a suitably small perturbation, $\lVert\delta\mathbf{x}_0\rVert\ll1$, 
\begin{equation}\label{eq:lyap_eq0}
\mathbf{x}\left(t\right)+\delta\mathbf{x}\left(t\right)=\Phi^t\left(\mathbf{x}_0\right)+D\Phi^t\left(\mathbf{x}_0\right)\delta\mathbf{x}_0+\mathcal{O}\left(\lVert\delta\mathbf{x}_0\rVert^2\right)\;\;\implies\;\;\delta\mathbf{x}\left(t\right)=D\Phi^t\left(\mathbf{x}_0\right)\delta\mathbf{x}_0
\end{equation}

\noindent where $D\Phi^t\left(\mathbf{x}_0\right)$ is the Jacobian matrix of $\Phi^t\left(\mathbf{x}_0\right)$, and
\begin{equation}\label{eq:lyap_eq1}
\lVert\delta\mathbf{x}\left(t\right)\rVert^2=\delta\mathbf{x}_0^\top\underbrace{\left(D\Phi^t\left(\mathbf{x}_0\right)\right)^\top D\Phi^t\left(\mathbf{x}_0\right)}_{\equiv\mathbf{M}^t\left(\mathbf{x}_0\right)}\delta\mathbf{x}_0
\end{equation}

\noindent where we have defined $\mathbf{M}^t\left(\mathbf{x}_0\right)$, the so-called Oseledets matrix. The matrix $\mathbf{M}^t\left(\mathbf{x}_0\right)$ is Hermitian and positive definite; therefore, its eigenvectors, which we denote $\mathbf{u}_k\left(t,\mathbf{x}_0\right)$ with $k=1,\ldots,N$, form a complete orthonormal basis and its eigenvalues, which we denote $\tilde{\Lambda}_k\left(t,\mathbf{x}_0\right)$, are all real and positive. We follow the time-honored tradition of ordering the eigenvalues in decreasing order, i.e., $\tilde{\Lambda}_1\left(t,\mathbf{x}_0\right)\geq\tilde{\Lambda}_2\left(t,\mathbf{x}_0\right)\geq\ldots\geq\tilde{\Lambda}_N\left(t,\mathbf{x}_0\right)$ for any $t$ and $\mathbf{x}_0$. Note that, however, the $\mathbf{u}_k\left(t,\mathbf{x}_0\right)$ change depending on both $t$ and $\mathbf{x}_0$.

We see from Eq.~\eqref{eq:lyap_eq1} that the component of the initial perturbation along $\mathbf{u}_k\left(t,\mathbf{x}_0\right)$ is multiplied by $\sqrt{\tilde{\Lambda}_k\left(t,\mathbf{x}_0\right)}$ as a result of the time evolution; thus, after a time $t$, that component is amplified if $\tilde{\Lambda}_k\left(t,\mathbf{x}_0\right)>1$, while it is suppressed if $0<\tilde{\Lambda}_k\left(t,\mathbf{x}_0\right)<1$. If this multiplicative amplification/suppression of the perturbation were to occur at a constant rate along the trajectory (i.e., exponentially with time), the rate would be 
\begin{equation}
    \frac{1}{t}\log\sqrt{\tilde{\Lambda}_k\left(t,\mathbf{x}_0\right)}=\frac{1}{2t}\log\tilde{\Lambda}_k\left(t,\mathbf{x}_0\right)
\end{equation}

\noindent The corresponding LE, $\Lambda_k\left(\mathbf{x}_0\right)$, is simply the limit for $t\to+\infty$ of the above expression, i.e., 
\begin{equation}
    \label{eq:lyap_skew}\Lambda_k\left(\mathbf{x}_0\right)\equiv\lim_{t\to+\infty}\frac{1}{2t}\log\tilde{\Lambda}_k\left(t,\mathbf{x}_0\right)
\end{equation}

\noindent if it exists. The celebrated Oseledets multiplicative ergodic theorem guarantees that, if the dynamics is ergodic, then (i) the above limit exists for almost any choice of $\mathbf{x}_0$; and that (ii) its value is independent of $\mathbf{x}_0$ (see, e.g., \cite{eckmann1985ergodic}). In general, ergodicity cannot be really verified {\em a priori} for many dynamical systems of practical interest (e.g., our random model neuronal network). In practice, one estimates the $\Lambda_k$, by numerically integrating the dynamical equations, and hopes for the best; indeed, even the numerical estimation is far from trivial (see Section \ref{sec:lyap_numerics} below, and \cite{engelken2023lyapunov}). 

We now express the Oseledets matrix in terms of the Jacobian matrix of $\mathbf{v}_\mathbf{w}\left(\mathbf{x}\right)$. To that end, we take the derivative with respect to time in Eq.~\eqref{eq:lyap_eq0}, and obtain
\begin{equation}\label{eq:lyap_eq2}
    \delta\dot{\mathbf{x}}=\mathbf{J}_\mathbf{w}\left(\mathbf{x}\left(t\right)\right)\delta\mathbf{x}
\end{equation}

\noindent where $\mathbf{J}_\mathbf{w}\left(\mathbf{x}\left(t\right)\right)$ is the Jacobian matrix of $\mathbf{v}_\mathbf{w}\left(\mathbf{x}\right)$ evaluated at $\mathbf{x}\left(t\right)$, i.e., the unperturbed trajectory at time $t$. The solution of Eq.~\eqref{eq:lyap_eq2}, with initial condition $\delta\mathbf{x}\left(0\right)=\delta\mathbf{x}_0$ is given by $\delta\mathbf{x}\left(t\right)=\mathbf{Y}\left(t\right)\delta\mathbf{x}_0$, where $\mathbf{Y}\left(t\right)$ is the fundamental matrix, i.e., the solution of
\begin{equation}\label{eq:lyap_eq3}
    \dot{\mathbf{Y}}=\mathbf{J}_\mathbf{w}\left(\mathbf{x}\left(t\right)\right)\mathbf{Y}
\end{equation}

\noindent with $\mathbf{Y}\left(0\right)=\mathbf{I}$ \cite{betounes_book}. Clearly, the Oseledets matrix can be expressed in terms of $
\mathbf{Y}\left(t\right)$ (see Eq.~\eqref{eq:lyap_eq1}). Henceforth, we do not explicitly indicate dependence on time (e.g., of $\mathbf{J}_\mathbf{w}$) to avoid notational clutter. 

Next, we assume that $\mathbf{J}_\mathbf{w}$ is diagonalizable along the trajectory. This assumption is numerically verified -- we find numerically that the spectrum of $\mathbf{J}_\mathbf{w}$ is self-averaging along the trajectory -- and theoretically motivated by random matrix theory \cite{ahmadian2015properties}, which predicts the spectrum of $\mathbf{J}_\mathbf{w}$ in our system (see Fig.~\ref{supfig4}(a)). Accordingly, for each time $t$, we can write
\begin{equation}\label{eq:lyap_eq4}
    \mathbf{J}_\mathbf{w}=\mathbf{V}\mathbf{D}\mathbf{V}^{-1};\;\;\;\;\mbox{with $D_{kl}=\lambda_k\delta_{kl}$}
\end{equation}

\noindent where $\mathbf{D}$ is a diagonal matrix that contains the instantaneous eigenvalues of $\mathbf{J}_\mathbf{w}$ and $\mathbf{V}$ is an $N\times N$ matrix whose $k$-th column is the eigenvector of $\mathbf{J}_\mathbf{w}$ associated to the $k$-th eigenvalues, $\lambda_k$. 

\noindent We study the linear dynamics of Eq.~\eqref{eq:lyap_eq3} in the time-varying basis formed by the eigenvectors $\mathbf{V}$. Accordingly, we define $\mathbf{X}$ such that $\mathbf{Y}=\mathbf{V}\mathbf{X}$. Its time evolution is given by
\begin{align}
    \dot{\mathbf{Y}} & =\mathbf{J}_\mathbf{w}\mathbf{Y}=\mathbf{J}_\mathbf{w}\mathbf{V}\mathbf{X}=\dot{\mathbf{V}}\mathbf{X}+\mathbf{V}\dot{\mathbf{X}} \\ & \implies \mathbf{V}^{-1}\mathbf{J}_\mathbf{w}\mathbf{V}\mathbf{X}=\mathbf{V}^{-1}\left(\dot{\mathbf{V}}\mathbf{X}+\mathbf{V}\dot{\mathbf{X}}\right) \\\label{eq:lyap_eq5} & \implies
    \dot{\mathbf{X}}=\left(\mathbf{D}-\underbrace{\mathbf{V}^{-1}\dot{\mathbf{V}}}_{\equiv\mathbf{A}}\right)\mathbf{X}
\end{align}

\noindent where we have defined the matrix $\mathbf{A}$. If $\mathbf{A}$ were the null matrix, then the LEs would simply reduce to the time-averaged real part of the eigenvalues, i.e., 
\begin{equation}\label{eq:lyap_eq6}
    \Lambda_i=\lim_{t\to\infty}\frac{1}{t}\int_{0}^{t}\mathrm{d}t'\,\operatorname{Re}\left(\lambda_i\left(t'\right)\right)\equiv\operatorname{Re}\left(\overline{\lambda_i}\right)
\end{equation}

\noindent where the overline denotes long-time averaging (i.e., along an infinitely long trajectory). In this case, we see from Eq.~\eqref{eq:lyap_eq6} that the leading LE cannot be larger than the maximum real part of the Jacobian spectrum along the trajectory. Analogously, the smallest LE cannot be smaller than the minimum real part of the Jacobian spectrum along the trajectory. 

In general, the effect of $\mathbf{A}$ cannot be ignored. To account for it, we perform a perturbative expansion of $\mathbf{X}$, the solution of Eq.~\eqref{eq:lyap_eq5}, as a power series in $\mathbf{A}$ (analogous to a Dyson series). That is,  $\mathbf{X}=\sum_{n=0}^\infty\mathbf{X}^{\left(n\right)}$, where the terms $\mathbf{X}^{\left(n\right)}$ satisfy the recursive relation
\begin{equation}\label{eq:lyap_eq_series}
\dot{\mathbf{X}}^{\left(0\right)}=\mathbf{D}\mathbf{X}^{\left(0\right)};\;\;\;\dot{\mathbf{X}}^{\left(n\right)}=\mathbf{D}\mathbf{X}^{\left(n\right)}-\mathbf{A}\mathbf{X}^{\left(n-1\right)}
\end{equation}

\noindent The zeroth-order solution, $\mathbf{X}^{\left(0\right)}$, is given by
\begin{equation}
X^{\left(0\right)}_{ij}\left(t\right)=\delta_{ij}\exp\left(\int_{0}^{t}\mathrm{d}t'\,\lambda_i\left(t'\right)\right)
\end{equation}

\noindent where, for technical convenience, we have chosen the identity matrix as the initial condition for the zeroth-order solution (i.e., $\mathbf{X}^{\left(0\right)}\left(0\right)=\mathbf{I}$). Note that,  because we are only interested in the asymptotic behavior (i.e., $t\to\infty)$, the choice of the initial condition is immaterial provided the initial matrix is non-singular.

\noindent The first-order correction, $\mathbf{X}^{\left(1\right)}$, is given by 
\begin{equation}
X^{\left(1\right)}_{ij}\left(t\right)=X^{\left(0\right)}_{ii}\left(t\right)\int_{0}^{t}\mathrm{d}t'A_{ij}\left(t'\right)\exp\left(\int_{0}^{t'}\mathrm{d}t''\,\left(\lambda_j\left(t''\right)-\lambda_i\left(t''\right)\right)\right)
\end{equation}

\noindent There are three cases to consider: $\operatorname{Re}\left(\overline{\lambda_j}\right)>\operatorname{Re}\left(\overline{\lambda_i}\right)$, $\operatorname{Re}\left(\overline{\lambda_j}\right)<\operatorname{Re}\left(\overline{\lambda_i}\right)$ and $\operatorname{Re}\left(\overline{\lambda_j}\right)=\operatorname{Re}\left(\overline{\lambda_i}\right)$. We refer to the first two-cases as non-resonant and to the last case as resonant (on average). Let us start with $\operatorname{Re}\left(\overline{\lambda_j}\right)>\operatorname{Re}\left(\overline{\lambda_i}\right)$. In this case, for large $t$
\begin{equation}
X^{\left(1\right)}_{ij}\left(t\right)\sim X^{\left(0\right)}_{ii}\left(t\right)A_{ij}\left(t\right)\exp\left(\int_{0}^{t}\mathrm{d}t'\,\left(\lambda_j\left(t'\right)-\lambda_i\left(t'\right)\right)\right)\sim A_{ij}\left(t\right)\exp\left(\int_{0}^{t}\mathrm{d}t'\,\lambda_j\left(t'\right)\right)
\end{equation}

\noindent where we use $\sim$ to denote equality to leading exponential order. If $A_{ij}\left(t\right)$ is bounded along the trajectory, i.e., $\lvert A_{ij}\left(t\right)\rvert\leq C$ where $C$ is independent of time, then these terms grow as $\exp\left(t\operatorname{Re}\left(\overline{\lambda_j}\right)\right)$ to leading exponential order. Next, we consider $\operatorname{Re}\left(\overline{\lambda_j}\right)<\operatorname{Re}\left(\overline{\lambda_i}\right)$. In this case, 
\begin{equation}
X^{\left(1\right)}_{ij}\left(t\right)\sim X^{\left(0\right)}_{ii}\left(t\right)\overline{A_{ij}}\sim \overline{A_{ij}}\exp\left(\int_{0}^{t}\mathrm{d}t'\,\lambda_i\left(t'\right)\right)
\end{equation}

\noindent where $\overline{A_{ij}}$ denotes the average of $A_{ij}\left(t\right)$ along the trajectory; $\overline{A_{ij}}$ exists, under the assumption that $A_{ij}\left(t\right)$ is bounded (almost everywhere) along the trajectory. Hence, these terms grow as $\exp\left(t\operatorname{Re}\left(\overline{\lambda_i}\right)\right)$ to leading exponential order.

In summary, the prefactors of non-resonant terms can be uniformly bounded by time-independent constants. This holds for all orders, $n$, of the expansion, as can be checked by repeating the same analysis for $n\geq 2$ (using Eq.~\eqref{eq:lyap_eq_series}). In turn, this ensures that no additional time-dependent growth beyond the exponential rates set by $\mathbf{D}$ can appear when re-summing the series.

Finally, we turn to $\operatorname{Re}\left(\overline{\lambda_j}\right)=\operatorname{Re}\left(\overline{\lambda_i}\right)$. In this case,
\begin{equation}
X^{\left(1\right)}_{ij}\left(t\right)\sim X^{\left(0\right)}_{ii}\left(t\right)\int_{0}^{t}\mathrm{d}t'A_{ij}\left(t'\right)\sim t\overline{A_{ij}}\exp\left(\int_{0}^{t}\mathrm{d}t'\,\lambda_i\left(t'\right)\right)
\end{equation}

\noindent and, hence, the prefactor of the resonant terms cannot be uniformly bounded by time-independent constants. Again, this holds for all orders, $n$, of the expansion; indeed, one can see that the prefactor of the resonant terms at order $n$ is proportional to $t^n$. When re-summing the series, this would lead to finite corrections to the exponential rates set by $\mathbf{D}$. However, if $\overline{A_{ij}}=0$ and the auto- and cross-correlations between the elements of $\mathbf{A}$ -- that is, $\overline{A_{ij}\left(t\right)A_{kl}\left(t'\right)}$ -- decay sufficiently fast along the trajectory so that the prefactors of the resonant terms only grow as $t^{n/2}$ at most, then the resulting corrections to the exponential rates become negligible.

In summary, our heuristic analysis shows that the leading exponential growth of the norm of $\mathbf{X}$ (determined by $\mathbf{D}$) remains unaffected by the presence of $\mathbf{A}$ if: (i) $A_{ij}\left(t\right)$ is bounded almost everywhere on the attractor; (ii) its time average vanishes ($\overline{A_{ij}}=0$); and (iii) its temporal correlations, $\overline{A_{ij}\left(t\right)A_{kl}\left(t'\right)}$, decay sufficiently fast for $\lvert t-t'\rvert\to\infty$. Physically, we expect these properties of $\mathbf{A}$ to be a direct consequence of the underlying chaotic dynamics, i.e., extensive chaos. The irregular nature of the trajectory drives a rapid decorrelation and an 'unbiased' re-orientation of the moving eigenvector basis, which is sufficient to suppress the growth of resonant terms so that they do not contribute to the leading exponential order. Thus, the Lyapunov exponents are well-approximated by the time-averaged real parts of the Jacobian spectrum, as observed in Fig.~\ref{supfig4}. At the same time, $\mathbf{A}$ retains enough structure to prevent the Lyapunov spectrum from collapsing to a single degenerate value. Because $\mathbf{A}=\mathbf{V}^{-1}\dot{\mathbf{V}}$ is constrained by the geometric structure of the attractor, its fluctuations cannot be entirely `isotropic'. In this sense, the statistical properties of the coupling matrix $\mathbf{A}$ encode meaningful information about the geometry of the chaotic attractor.

\section{Numerical simulations}
\subsection{Newton's method with backtracking}
We use Newton's method with a backtracking line search to find equilibria of the dynamics. At each iteration, the update is computed as
\begin{equation}
    \Delta\mathbf{x} = -\mathbf{J}^{-1}(\mathbf{x})\big(\sqrt{N}-\mathbf{x}+\frac{1}{\sqrt{N}}\mathbf{w}\mathbf{\phi}\big), \quad 
    \mathbf{x} \to \mathbf{x} + \alpha\Delta\mathbf{x},
\end{equation}
where $\mathbf{J}(\mathbf{x})$ is the Jacobian of the dynamical equation, and $\mathbf{\phi}=\phi(\mathbf{x})$ is the nonlinear activation. The update is accepted if it decreases the objective function, defined as the norm of the velocity, $f(\mathbf{x}) = |\sqrt{N}-\mathbf{x}+\frac{1}{\sqrt{N}}\mathbf{w}\mathbf{\phi}|$. Otherwise, the step size $\alpha$ (initialized as $1$) is reduced by a factor of $0.5$ until the condition below is satisfied
\begin{equation}
    f(\mathbf{x}+\alpha\Delta\mathbf{x}) \le f(\mathbf{x}) + c(f(\mathbf{x}+\Delta\mathbf{x})-f(\mathbf{x})). 
\end{equation}
We set $c=0.001$ as the backtracking parameter and set the maximum number of iterations to $1000$. The algorithm stops when $f(\mathbf{x})$ is smaller than a predefined threshold, $f(\mathbf{x})<10^{-3}$, or when the maximum number of iterations is reached. For each network realization, we run Newton's method with $10^6$-$10^7$ different initial conditions, independently and uniformly sampled within an $N$-dimensional hypercube with large enough side length, i.e., $L=\sqrt{N}$. This choice ensures that the entire phase space is explored uniformly, without bias toward any particular region. 

\paragraph{Determining the number of initial conditions}
To reliably identify all distinct equilibria, we do not fix the number of initial conditions {\em a priori}. Instead, we choose it adaptively to ensure that each unique equilibrium (see below for the definition of uniqueness) is, on average, discovered at least $m$ times from independent searches. Larger values of $m$ improve the robustness of the complexity estimate, but also increase computational cost. In our numerical search, we set $m=30$ as a practical compromise between accuracy and efficiency. 

\paragraph{Identifying true and unique equilibria}
To compare the numerical results with theory, it is crucial to accurately count the number of true and unique equilibria identified by the algorithm. In practice, it is challenging to distinguish a true equilibrium from a slow point (a location where the velocity is small but nonzero), and to decide whether two nearby equilibria should be treated as identical within finite machine precision. This problem can be addressed by using threshold-linear activation units. In this case, each equilibrium is uniquely characterized by a binary configuration vector, $\mathbf{s}=\Theta(\mathbf{x})$, and the corresponding criterion can be written as
\begin{equation}
    (2s_i-1)h_i + \frac{1}{\sqrt{N}}\sum_{j=1}^Nw_{ij}s_jh_j = \sqrt{N},\quad i=1\dots N, 
\end{equation}
where $s_i=\Theta(x_i)$ and $h_i=|x_i|$. $\mathbf{x}$ is classified as a true equilibrium if the solution exists and all components lie in the positive orthant, i.e., $h_i\ge 0$ for all $i=1\dots N$.

\subsection{Numerical estimation of complexity}
\begin{figure}[h]
    \centering
    \includegraphics[width=0.95\linewidth]{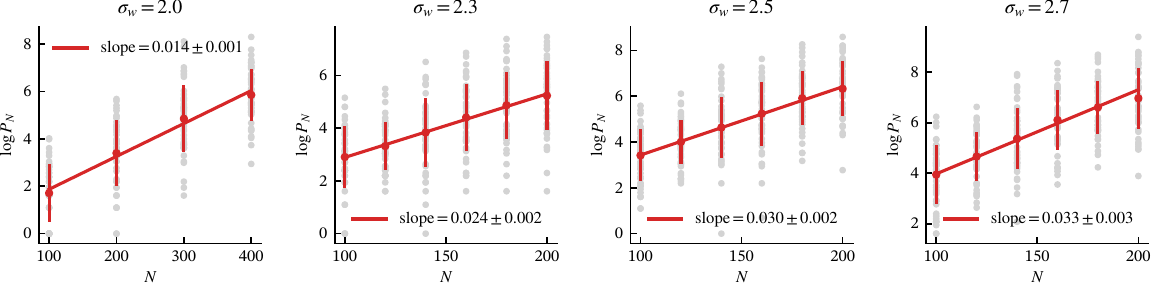}
    \caption{Numerical estimation of $\Sigma_q^*$. Grey dots: $\log P$ for $M=50$ sample networks. Error bars: mean and SD across sample networks. 
    }
    \label{supfig1}
\end{figure}

The algorithm allows us to compare theory with estimates of $\Sigma_q^*$ obtained by counting the number of equilibria in sample networks. For each pair $(\sigma_w, N)$, we generate $M=50$ network realizations and search for their equilibria using the algorithm described above. For fixed $\sigma_w$, we regress $\log P$ against $N$ (with $M$ samples of $\log P$ for each $N$) to obtain an estimate of $\Sigma_q^*$ at that $\sigma_w$. The results of this procedure are shown in Fig.~\ref{supfig1}. 

\subsection{Numerical estimation of order parameters}

Let us consider a given sample network and let $\boldsymbol{\phi}^a$ $(a=1,\ldots,P)$ be its equilibria. We define $S\left(n\right)$, with $n=0\ldots,N$, the set of the indexes of the equilibria which have exactly $n$ neurons active, i.e.,

\begin{equation}
    S\left(n\right)\equiv\left\{a\;:\;\sum_i\Theta\left(x^a_i\right)=n,\;1\leq a\leq P\right\}
\end{equation}

\noindent Note that there will be $n$ for which $S\left(n\right)=\emptyset$, i.e., $S\left(n\right)$ is the empty set. The upcoming analysis is restricted to values of $n$ for which $S(n)$ contains at least one element. Clearly,
\begin{equation}
    \bigcup_{n=0}^NS\left(n\right)=\left\{1,\ldots,P\right\};\;\;\;S\left(n_1\right)\cap S\left(n_2\right)=\emptyset\;\;\;\mbox{if $n_1\neq n_2$}
\end{equation}

\noindent We denote $P\left(n\right)$ the number of elements in $S\left(n\right)$, i.e., its cardinality, $P\left(n\right)\equiv\lvert S\left(n\right)\rvert$. $P\left(n\right)=0$ if $S\left(n\right)$ is empty. Clearly,

\begin{equation}
    \sum_{n=0}^NP\left(n\right)=P
\end{equation}

\noindent We consider values of $n$ for which $S\left(n\right)$ is not empty. For these $n$, we compute

\begin{equation}
    Q^a\left(n\right)=\frac{1}{N}\sum_i\left(\phi^a_i\right)^2;\;\;a\in S\left(n\right)
\end{equation}

\noindent From these $P\left(n\right)$ quantities, we compute $\hat{Q}(n)$ as 
\begin{equation}
    \hat{Q}(n)=\frac{1}{P\left(n\right)}\sum_{a\in S\left(n\right)}Q^a(n)
\end{equation}

\noindent Thus, $\hat{Q}\left(n\right)$ is a tabulated estimate of $Q\left(f\right)$ $\left(n=fN\right)$ in a {\em given} sample network.

We can similarly obtain a tabulated estimate of $q\left(f\right)$, $\hat{q}\left(n\right)$, in a given sample network. For this, we compute

\begin{equation}
    q^{ab}\left(n\right)=\frac{1}{N}\sum_i\phi^a_i\phi^b_i;\;\;a,b\in S\left(n\right),\;a<b
\end{equation}

\noindent There are $\frac{1}{2}P\left(n\right)\left(P\left(n\right)-1\right)$ such quantities, from which we compute

\begin{equation}
    \hat{q}\left(n\right)=\frac{2}{P\left(n\right)\left(P\left(n\right)-1\right)}\sum_{\substack{a,b\in S\left(n\right) \\ a<b}}q^{ab}\left(n\right)
\end{equation}

\noindent One can also obtain a tabulated estimate of the overlap between two equilibria with different $f$. For this one computes

\begin{equation}
    q^{ab}\left(n_1,n_2\right)=\frac{1}{N}\sum_i\phi^a_i\phi^b_i;\;\;a\in S\left(n_1\right),\; b\in S\left(n_2\right)
\end{equation}

\noindent There are $P\left(n_1\right)P\left(n_2\right)$ such quantities, from which one computes

\begin{equation}
    \label{eq: overlap_centroids}
    \hat{q}\left(n_1,n_2\right)=\frac{1}{P\left(n_1\right)P\left(n_2\right)}\sum_{\substack{a\in S\left(n_1\right) \\ b\in S\left(n_2\right)}}q^{ab}\left(n_1,n_2\right)
\end{equation}

\subsection{Numerical evaluation of the Lyapunov spectrum}
\label{sec:lyap_numerics}
The Lyapunov exponents of a continuous-time dynamical system can be computed using Benettin's algorithm \cite{benettinetal1980b}; see \cite{sandri1996numerical} for a pedagogical introduction. The method integrates the state dynamics together with the associated \emph{variational equation}, which describes the evolution of infinitesimal perturbations. 

\noindent Recall the dynamical equation
\begin{equation}
    \dot{x}_i = -x_i + \sqrt{N} + \frac{1}{\sqrt{N}}\sum_{j=1}^N w_{ij}\phi_j
    \equiv v_i(\mathbf{x}),
\end{equation}
where $\mathbf{v}(\mathbf{x})$ is the \emph{vector field}.
The \emph{flow} $\Phi^t:\mathbf{x}_0\mapsto\mathbf{x}(t)$ generated by
$\mathbf{v}$ satisfies
\begin{equation}
    \dot{\Phi}^t(\mathbf{x}_0) = \mathbf{v}(\Phi^t(\mathbf{x}_0)),\quad
    \Phi^0(\mathbf{x}_0) = \mathbf{x}_0.
\end{equation}

\noindent For two nearby initial conditions $\mathbf{x}_0$ and $\mathbf{x}_0+\delta\mathbf{x}_0$, the linearized separation evolves as 
\begin{equation}
    \delta\mathbf{x}(t)
    = \Phi^t(\mathbf{x}_0+\delta\mathbf{x}_0)-\Phi^t(\mathbf{x}_0)
    \approx \mathbf{U}^t(\mathbf{x}_0)\,\delta\mathbf{x}_0,
\end{equation}
where $\mathbf{U}^t(\mathbf{x}_0)\equiv D_{\mathbf{x}_0}\Phi^t(\mathbf{x}_0)$ is the \emph{tangent map}. Differentiating the flow equation with respect to $\mathbf{x}_0$ gives the variational equation
\begin{equation}
    \dot{\mathbf{U}}^t(\mathbf{x}_0)
    = \mathbf{J}(\Phi^t(\mathbf{x}_0))\cdot\mathbf{U}^t(\mathbf{x}_0),
    \quad \mathbf{U}^0(\mathbf{x}_0) = \mathbf{I}, 
\end{equation}
where $\mathbf{J}(\mathbf{x})=\partial_{\mathbf{x}}\mathbf{v}(\Phi^t(\mathbf{x}_0))$ is the instantaneous Jacobian
evaluated along the unperturbed trajectory. Thus the state and tangent dynamics are integrated jointly as 
\begin{equation}
    \begin{bmatrix}\dot{\mathbf{x}}\\\dot{\mathbf{U}}\end{bmatrix}
    =\begin{bmatrix}
        \mathbf{v}(\mathbf{x})\\
        \mathbf{J}(\mathbf{x})\cdot\mathbf{U}
    \end{bmatrix},
    \quad\mathbf{x}(0)=\mathbf{x}_0,\quad\mathbf{U}(0)=\mathbf{I}.
\end{equation}

\noindent By Oseledets' theorem, the Lyapunov exponents are obtained from the eigenvalues $\tilde{\Lambda}_k$ of $(\mathbf{U}^t)^T\mathbf{U}^t$:
\begin{equation}
    \Lambda_k
    = \lim_{t\to\infty}\frac{1}{2t}
      \log\tilde{\Lambda}_k\!\left((\mathbf{U}^t)^T\mathbf{U}^t\right).
\end{equation}

\noindent The coupled equations can be integrated straightforwardly by any ODE solver. For a chaotic system, however, the norm of $\mathbf{U}^t$ grows exponentially fast, causing numerical overflow. To avoid this instability, we follow Benettin's algorithm by periodically applying QR decomposition to $\mathbf{U}^t$. At the $s$-th re-orthonormalization step, the diagonal entries of the upper-triangular factor $\mathbf{R}^{(s)}$ are accumulated, leading to 
\begin{equation}
    \Lambda_k = \lim_{T\to\infty}
    \frac{1}{T}\sum_{s=1}^{T/\Delta t}\log\bigl\lvert \mathbf{R}_{kk}^{(s)}\bigr\rvert. 
\end{equation}

In practice, we compute only the leading $p\ll N$ exponents by initializing $\mathbf{U}^0 \in \mathrm{R}^{N\times p}$ as the first $p$ columns of the identity matrix. For the Lyapunov-dimension estimates in the main text, we track the top 10\% of exponents with $N=2000$, $T=5000$ and $\Delta t=0.1$. We verified \emph{a posteriori} that, for all simulations used in this estimate, the cumulative sum of the computed Lyapunov exponents crossed zero within the tracked portion of the spectrum. Full spectra are computationally expensive and are therefore computed only for smaller networks with $N=500$ and $T=2000$. 

\begin{tcolorbox}[float=h, colback=white, colframe=black, boxrule=0.4pt, arc=0pt]
    \begin{algorithmic}[1] 
    \State \textbf{Input:} weight matrix $\mathbf{W}$, time $T$, step $\Delta t$, number of tracked exponents $p$
    \State Initialize $\mathbf{x}(0)\sim\mathcal{N}(0,\mathbf{I})$,\quad $\mathbf{U}^0 = \mathbf{I}_N[:, {:}p]$,\quad $\mathbf{\gamma}=\mathbf{0}\in\mathbb{R}^p$
    \For{$s = 1,\ldots, T/\Delta t$}
        \State $(\mathbf{x},\mathbf{U})\leftarrow\mathrm{RK4}\bigl([\dot{\mathbf{x}}=\mathbf{v}(\mathbf{x}),\; \dot{\mathbf{U}}=\mathbf{J}(\mathbf{x})\mathbf{U}],\;\Delta t\bigr)$
        \hfill where $\mathbf{J}(\mathbf{x})=-\mathbf{I}+\frac{1}{\sqrt{N}}\mathbf{W}\,\mathrm{diag}(\Theta(\mathbf{x}))$
        \State $\mathbf{Q},\mathbf{R}\leftarrow\mathrm{QR}(\mathbf{U})$
        \State $\mathbf{U}\leftarrow\mathbf{Q}$ \hfill (re-orthonormalize)
        \State $\mathbf{\gamma}\leftarrow\mathbf{\gamma}+\log\lvert\mathrm{diag}(\mathbf{R})\rvert$
    \EndFor
    \State Return $\hat{\Lambda}_k \leftarrow \gamma_k/T$ \quad for $k=1,\ldots,p$
    \end{algorithmic}
    \vspace{4pt}\centering{\textbf{Algorithm 1.} Benettin's algorithm for the top-$p$ Lyapunov exponents.}
\end{tcolorbox}

\bibliographystyle{unsrt}
\bibliography{refs.bib}

\clearpage
\section*{Supporting figures}

\begin{figure}[h]
    \centering
    \includegraphics[width=0.95\textwidth]{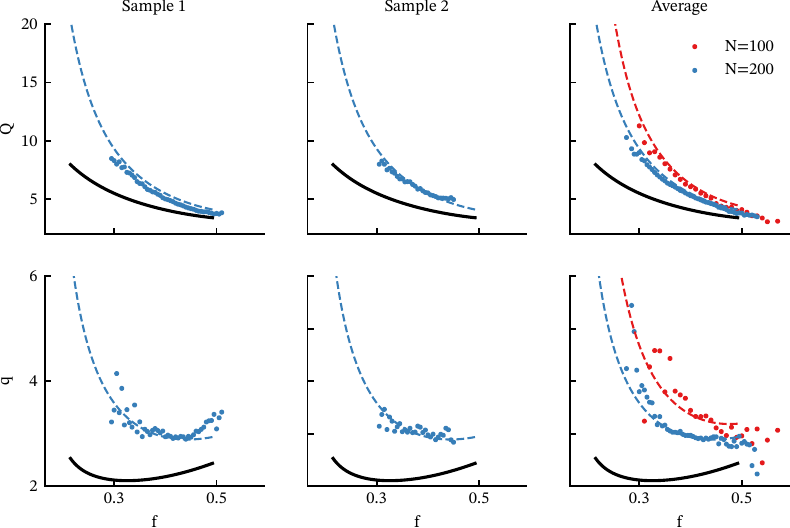}
    \caption{Estimated order parameters $Q(f)$ and $q(f)$ of two randomly selected sample networks (first two columns) and averaged over 50 sample networks (last column) for $\sigma_w=2.5$. Full black lines: asymptotic theory; Dashed lines: finite-$N$ theory; Dots: numerical simulations. }
    \label{supfig2}
\end{figure}

\begin{figure}[tb]
    \centering
    \includegraphics[width=0.95\textwidth]{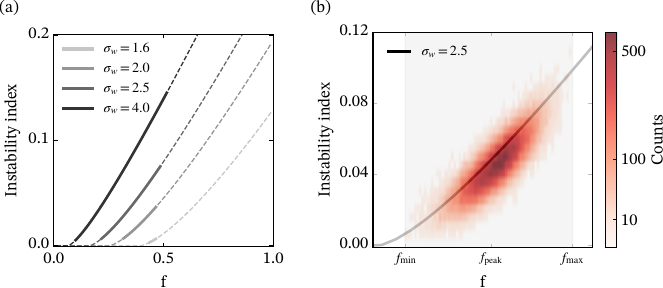}
    \caption{Comparison of theoretical and numerical instability indices. (a) Theoretical instability index $\alpha(f)$ versus $f$ for varying $\sigma_w$, derived from Eq.~\eqref{eq: instability-index}. Solid line: $f$ with positive complexity. (b) Joint distribution of $f$ and the instability index for $51,575$ numerical equilibria across $50$ networks ($\sigma_w=2.5$, $N=200$). Vertical markings  $f_{\min}$, $f_{\mathrm{peak}}$ and $f_{\max}$ indicate the boundaries and peak location predicted by the theory. }
    \label{supfig3}
\end{figure}

